\documentclass[sigconf]{acmart}
\usepackage[disable]{todonotes}
\usepackage{enumerate}
\usepackage[shortlabels]{enumitem}
\usepackage{amsmath}

\usepackage[inkscapelatex=false]{svg}
\usepackage{soul}

\usepackage{multirow}

\newcommand{\artem}[1]{}
\newcommand{\alex}[1]{}
\newcommand{\mathieu}[1]{}
\newcommand{\jl}[1]{}
\newcommand{\rebuttaldelete}[1]{}
\newcommand{\rebuttaldeleteeq}[1]{}
\newif\ifeq

\eqfalse

\AtBeginDocument{%
  \providecommand\BibTeX{{%
    \normalfont B\kern-0.5em{\scshape i\kern-0.25em b}\kern-0.8em\TeX}}}

\copyrightyear{2025}
\acmYear{2025}
\setcopyright{cc}
\setcctype{by}
\acmConference[CHI '25]{CHI Conference on Human Factors in Computing
Systems}{April 26-May 1, 2025}{Yokohama, Japan}
\acmBooktitle{CHI Conference on Human Factors in Computing Systems (CHI
'25), April 26-May 1, 2025, Yokohama,
Japan}\acmDOI{10.1145/3706598.3713631}

\begin{document}

%\title[]{LiveLocalizer: Localization and beamforming on low-power wearable devices using microphone array }
%\title[]{LiveLocalizer: Augmenting Mobile Speech-to-Text with Microphone Arrays and Optimized Localization}
%\title[]{SpeechCompass: Implementation and Evaluation of Speaker Separation Technology for Mobile Speech-to-Text Applications}
%\title[]{SpeechCompass: Implementation and Evaluation of Multimicrophone Diarization and Localization Technology for Mobile Speech-to-Text}
%\title[]{SpeechCompass: Enabling Multi-Speaker Mobile Captioning User Interfaces through Multimicrophone Diarization and Speech Localization}

%\title[]{SpeechCompass: Enabling Mobile Captioning for Multi-Speaker Conversations through Multi-microphone Diarization and Speech Localization}

%\title[]{SpeechCompass: Enabling Mobile Captioning for Multi-Speaker Conversations through Multi-microphone Speech Localization}

%\title[]{SpeechCompass: Diarized Captioning for Multi-Speaker Conversations through Multi-microphone Speech Localization on Mobile Devices}

%\title[]{SpeechCompass: Diarized Captioning on Mobile Devices for Multi-Speaker Conversations through Multi-microphone Speech Localization}

%\title[]{SpeechCompass: Diarized Mobile Captioning  for Group Conversations and Awareness through Multi-microphone Speech Localization}

\title[]{SpeechCompass: Enhancing Mobile Captioning with Diarization and Directional Guidance via Multi-Microphone Localization}


%\author{anonymous}
% \affiliation{%
%   \institution{X}
%   \city{X}
%   \country{X}}
% \email{X@X.com}

\author{Artem Dementyev*}
\affiliation{%
  \institution{Google Research}
  \city{Mountain View}
  \country{CA}}
\email{artemd@google.com}

\author{Dimitri Kanevsky}
\affiliation{%
  \institution{Google Research}
  \city{Mountain View}
  \country{CA}}
\email{dkanevsky@google.com}

\author{Samuel J. Yang}
\affiliation{%
  \institution{Google Research}
  \city{Mountain View}
  \country{CA}}
\email{samuely@google.com}

\author{Mathieu Parvaix}
\affiliation{%
  \institution{Google Research}
  \city{Mountain View}
  \country{CA}}
\email{parvaix@google.com}

\author{Chiong Lai}
\affiliation{%
  \institution{Google Research}
  \city{Mountain View}
  \country{CA}}
\email{chionglai@google.com }

\author{Alex Olwal*}
\affiliation{%
  \institution{Google Research}
  \city{Mountain View}
  \country{CA}}
\email{olwal@acm.org}

\thanks{*First and last author contributed equally to this work.} 

%\renewcommand{\shortauthors}{Anonymous et al.}

\begin{abstract}
Speech-to-text capabilities on mobile devices have proven helpful for hearing and speech accessibility, language translation, note-taking, and meeting transcripts. However, our foundational large-scale survey (n=263) shows that the inability to distinguish and indicate speaker direction makes them challenging in group conversations. SpeechCompass addresses this limitation through real-time, multi-microphone speech localization, where the direction of speech allows visual separation and guidance (e.g., arrows) in the user interface. We introduce efficient real-time audio localization algorithms and custom sound perception hardware, running on a low-power microcontroller with four integrated microphones, which we characterize in technical evaluations. Informed by a large-scale survey (n=494), we conducted an in-person study of group conversations with eight frequent users of mobile speech-to-text, who provided feedback on five visualization styles. The value of diarization and visualizing localization was consistent across participants, with everyone agreeing on the value and potential of directional guidance for group conversations. 
\end{abstract}



%\ccsdesc[500]{Human-centered computing~Graphics input devices}
%\ccsdesc[300]{Human-centered computing~Accessibility technologies; HCI; Mobile Computing; User Studies; Hardware}


\begin{teaserfigure}
  \centering
  \includegraphics[width=0.95\textwidth]{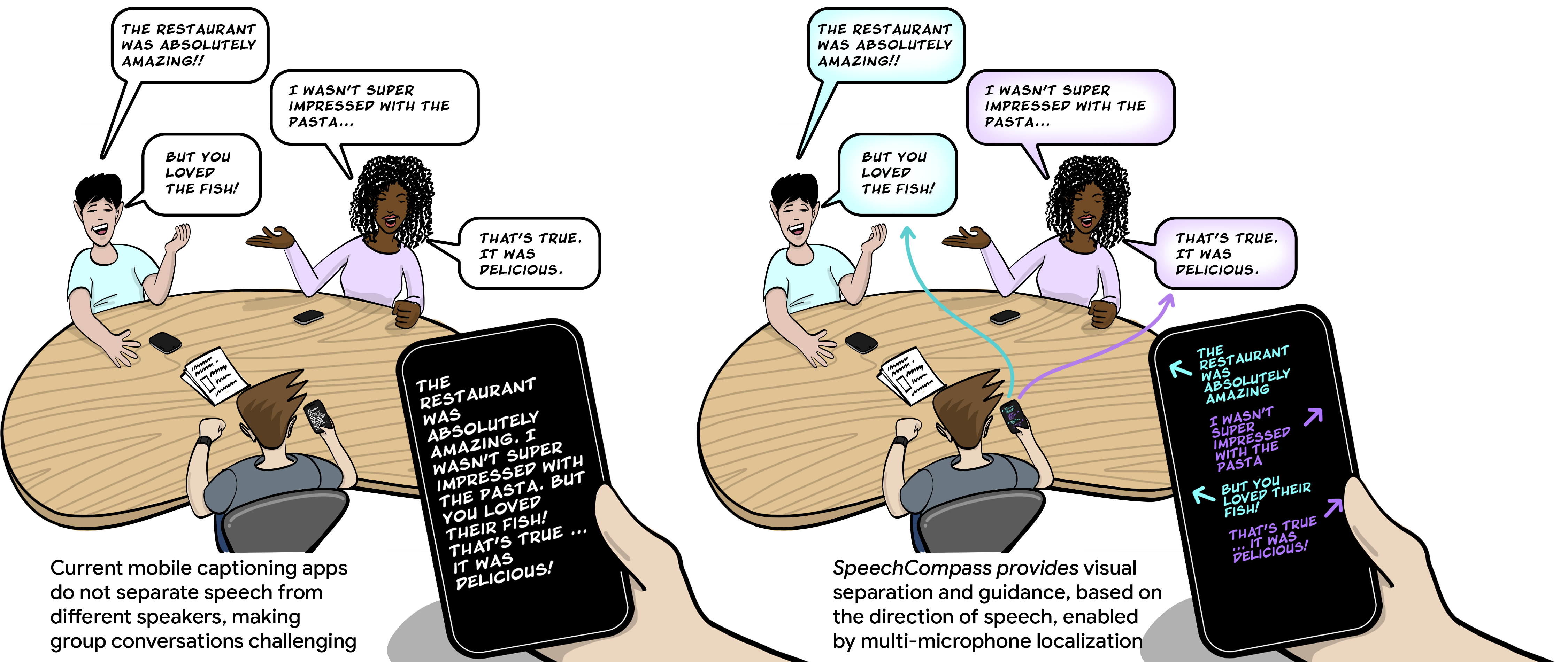}
    % \includegraphics[width=\textwidth/3]{images/illustrations/Panel 1.jpg}
    % \includegraphics[width=\textwidth/3]{images/illustrations/Panel 2_clean.png}
    % \includegraphics[width=0.3\textwidth]{images/illustrations/Panel 2 - Phone Close-Up.png}    
%    \includegraphics[width=\textwidth]{images/teaser.pdf}
%  \includegraphics[width=0.7\textwidth]{images/hidden_interfaces_concept_sketch.png}
%  \includegraphics[width=0.7\textwidth]{images/teaser_hidden_interfaces.pdf}
 % \caption{\emph{Subsurface interactions} introduce rectilinear user interface elements that can be rendered with high-brightness. They can thus be hidden behind textile, wood or plastic, and appear on demand. Left: An implementation of toggle selection controls. Right: Manipulation of continuous parameters using sliders.}
  \caption{\emph{SpeechCompass} creates user-friendly speech transcripts for group conversations with multiple speakers. \textit{Left}: Current solutions concatenate and mix the transcribed speech when multiple people participate in a conversation, which makes it challenging to read and understand the transcript. \textit{Right}: SpeechCompass addresses this limitation through real-time, multi-microphone speech localization, where \emph{the direction of speech} allows diarization, visual separation, and guidance (e.g., arrows) in the user interface. }
  \Description{}
  \label{fig:teasersketch}
\end{teaserfigure}

%\footnote{* "Authors contributed equally to this work."}

%\todo{add a, b, c, d, e, f, g to teaser figure}
\maketitle

\section{Introduction}
Recent advances have enabled real-time automatic speech recognition (ASR) in mobile and embedded hardware, supporting a range of conversational scenarios~\cite{LiveTranscribe, TranslatorMicrosoft}. Real-time captioning can enhance human communication in various ways, e.g., for real-time translation between languages, transcriptions during an interview, automatic subtitle generation, hearing accessibility~\cite{designing_asr_for_deaf, tracked_asr_study}, and note-taking. 
%%As well as provide a warning about a loud incoming sound. samuely@: I couldn't figure out how to work this back in.
However, there are still unsolved limitations with real-time speech-to-text --- specifically, the ability to distinguish multiple speakers (\textit{speaker diarization}) and tracking the direction of speech (\textit{localization}).

\begin{figure*}
  \centering
  \includegraphics[width=0.9\linewidth]{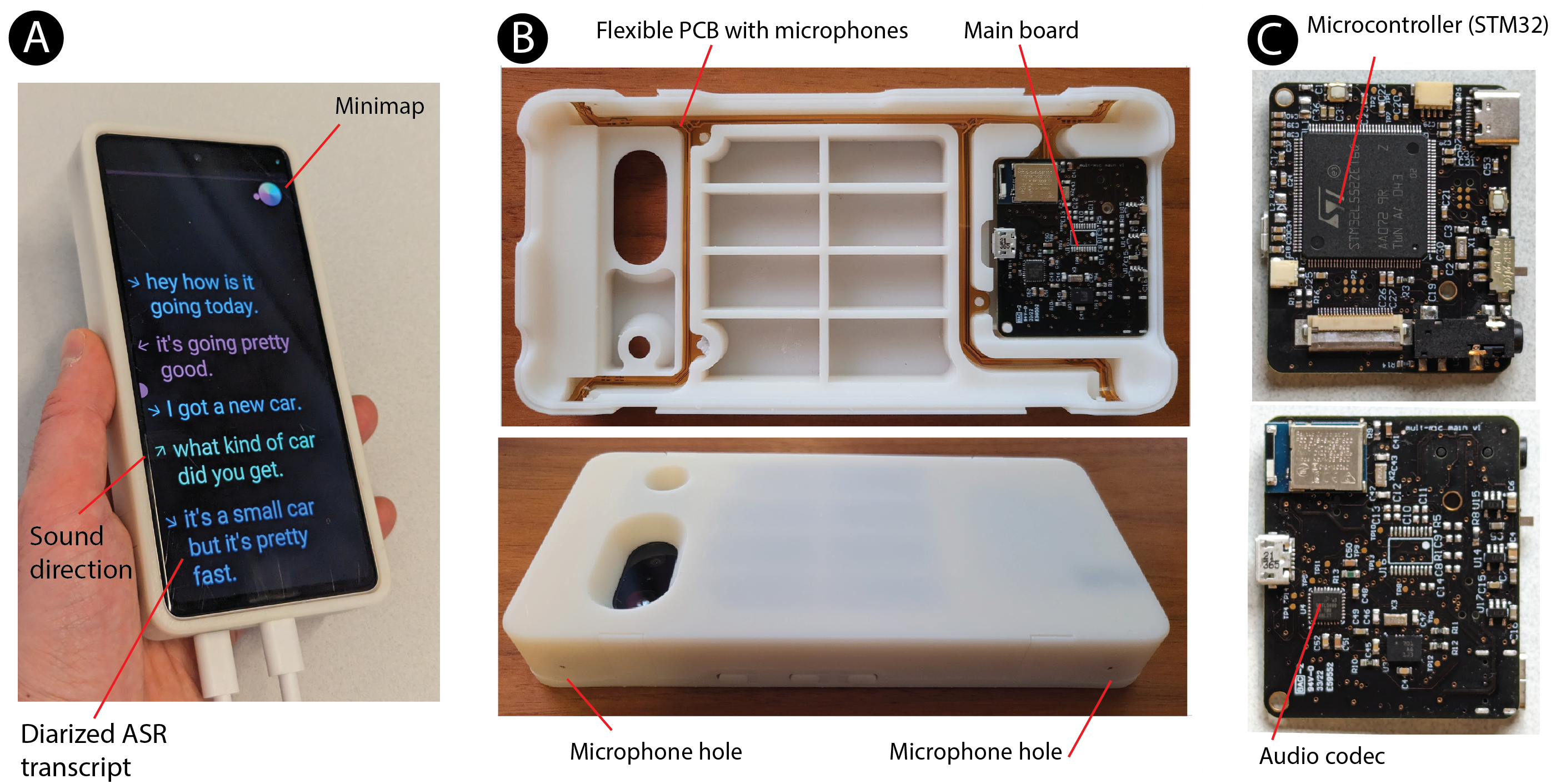}
  \caption{Overview of the SpeechCompass phone case prototype. A) A mobile phone application interface with a mounted multi-microphone phone case. B) Inside and outside view of the prototype with a flexible PCB microphone mount and a compact main printed circuit board (PCB). C) Pictures of the main PCB with a top and bottom view. }
  \label{fig:pcb_design}  
\end{figure*}

Consider the following scenario: 
\textit{Throughout the day, a person relies on a mobile phone with real-time captioning to understand speech.
However, at a dinner table with multiple people, the conversation is difficult to follow since the app cannot distinguish between speakers and concatenates all speech into a single paragraph.
Additionally, since the person needs to look at their phone regularly, they struggle with following the turn-taking across speakers and don't know where to look if there is a speaker change. Also, irrelevant nearby conversations get transcribed as well, which can cause confusion and privacy implications.  }

Many of the challenges in the scenario stem from the spatial complexities of audio, which are challenging to capture with a single-microphone ASR system. The benefits of multi-microphone topologies for localization have been demonstrated in numerous applications, such as public safety~\cite{valenzise2007scream}, virtual reality~\cite{vr_sound_localization}, robot navigation~\cite{liu2010continuous}, mobile computing~\cite{phone_localization, phone_localization2} and audio accessibility~\cite{localization_glasses,holo_sound}.
In this work, we leverage arrays of multiple microphones and apply techniques for microphone array signal processing to demonstrate how this technology could improve ASR performance and usability in such scenarios. Specifically, we identified opportunities for improvements in three areas:
\begin{enumerate}
    \item \textit{Speaker diarization}. The transcript can visually separate different speakers based on the direction of the speech. \item \textit{Localization}. For spatial sound visualization, the screen can display the direction of the incoming sound.
    \item \textit{Selective attention}. The interface can allow the user to select speech of interest and filter out self-speech.
\end{enumerate}

%Microphone array processing has traditionally been limited to high-end audio applications, and was too computationally intensive to meet the latency and power consumption requirements for mobile and wearable devices. Recent advances have, however, enabled multi-microphone algorithms to be used in low-power devices with low latency. 

In this work, we developed \textit{SpeechCompass}, a solution to add diarization and speech localization to mobile captioning. It includes three main parts. First, low-latency localization algorithms that can run on generic microcontrollers or mobile phones. Second, a compact 4-microphone phone case that allows 360-degree localization on a low-power microcontroller. \rebuttaldelete{We used the STM32 microcontroller, a popular professional choice but also available for hobbyists as it is used in many Arduino-compatible boards (e.g., Adafruit Feather Express)} Third, a mobile captioning app that shows how sound localization can be visualized in different ways and used to support multi-speaker conversations through diarized transcripts. We also run our algorithm on an off-the-shelf mobile phone with only two microphones to demonstrate that limited 180-degree localization is possible in the app without additional hardware.

While machine learning approaches to single-source speaker diarization have been improving~\cite{park2022review}, our multi-mic approach has the advantage of lower computational cost, latency, and privacy, and thus is inherently suitable for real-time ASR applications on low-power, low-cost microcontrollers. It is also language agnostic and can work for sounds other than speech. Our approach is tied to the position of the phone and the speaker, which can be advantageous as diarization can be immediately reconfigured by moving the phone. This paper shows how traditional microphone array processing can significantly benefit diarization and localization for mobile ASR. While diarization with microphone arrays has been well studied~\cite{anguera2007acoustic}, it has yet to be applied to mobile ASR and related interfaces.
%The use of our phone-case approach would enable this functionality to work for any phone, not just those with multiple built-in microphones. 
%However, this approach needs further characterization for scenarios with multiple \textit{simultaneous} speakers.
\jl{Tie this back better to UI, give CHI}

%While diarization has advanced significantly using machine learning~\cite{erdogan2015phase, isik2016single}, it is difficult to apply to mobile real-time diarization due to high compute and non-causal algorithms, as outlined in Table~\ref{table:tech_comparison}. Furthermore, the computational requirements for the models are large, so running privacy-preserving models on the device is challenging. This paper shows that diarization and mobile ASR can significantly benefit from traditional microphone array processing. Diarization with microphone arrays has been well studied~\cite{anguera2007acoustic} but has yet to be applied to mobile ASR and interactions.

\subsection{Contributions}
The contributions of this work are: 
\begin{itemize}
    \item \textbf{Real-time sound localization algorithms and optimized, embedded multi-microphone hardware}, implemented both as phone-case prototypes with a low-power microcontroller (<20 ms processing time) and on a mobile phone with constrained microphone hardware. The code and design files are available via Github.\footnote{\href{https://github.com/google/multi_mic_audio}{github.com/google/multi\_mic\_audio}}
    
     \item \textbf{Mobile captioning UIs for group conversations, enabled through speech localization and diarization}, and implemented as mobile speech-to-text applications with different visualization and interaction techniques.
        
    \item \textbf{Technical evaluation of localization and diarization using the optimized algorithms and hardware}. We characterize localization accuracy and estimation time under various signal-to-noise conditions and speaker configurations, and evaluate diarization accuracy.

    \item \textbf{Foundational large-scale survey (n=263) with frequent users of captioning technology}, showing that noise and speaker separation are important and frequent challenges with existing solutions.

    \item \textbf{Lab study (n=8) and large-scale survey (n=494) of mobile interfaces and visualizations with frequent users of captioning technology}. Informed by a large-scale survey, we designed a lab study with eight frequent users of captioning technology. We contribute insights into interface preferences, customization, and benefits since diarization and localization have not been previously studied for mobile ASR.
    
\end{itemize}

\section{Related work}
This section first discusses previous works in relevant audio applications like diarization, real-time audio transcription, and visualization interfaces. Then, we discuss the relevant literature on audio multi-microphone processing.

\subsection{Real-time mobile speech recognition}
CART (Communication Access Realtime Translation) is a well-established method for providing real-time captions.  A trained professional, typically using specialized software and a stenography machine, transcribes speech into text as it is spoken. CART is frequently used in broadcasting scenarios such as lectures and presentations. Recent advances in machine learning enabled real-time automatic mobile speech captioning, such as Live Transcribe~\cite{LiveTranscribe} and Microsoft Translator~\cite{TranslatorMicrosoft}, and also on head-worn displays~\cite{wearable_subtitle}. One of the main motivating uses for audio-to-text translation was audio accessibility, and research has shown the benefits of real-time ASR for hearing accessibility in scenarios like classrooms~\cite{asr_deaf_classroom, tracked_asr_study}. Research has shown various improvements to real-time ASR, such as ways to display transcription confidence~\cite{perspective_on_imperfect_captions} and customizing fonts appearance to improve readability~\cite{preffered_captions_appearance}.
\jl{Add more information regarding how we build on this related work}

\subsection{Speaker separation and diarization: Transcripts that distinguish speakers} 
% \subsection{Speaker separation and diarization: Enabling transcripts that distinguish between different speakers} %\alex{Seems like we could use this additional subsection heading?}\artem{Right, that makes it more clear}

Studies suggest that there is still room for improvements in, e.g., accuracy and usability~\cite{asr_challanges_deaf}. We particularly observe that existing approaches have yet to leverage microphone arrays in mobile devices to augment ASR. We outline the existing techniques in Table~\ref{table:tech_comparison}.

\begin{table*}
  \centering
\caption{Comparison of speaker diarization and separation technologies. The comparison shows that only SpeechCompass can support diarization and visualize sound direction on mobile devices. \jl{Can you not do acoustic beamforming on mobile? Is that the only difference from SpeechCompass?}}
  \label{table:tech_comparison}  
  \includegraphics[width=0.9\linewidth]{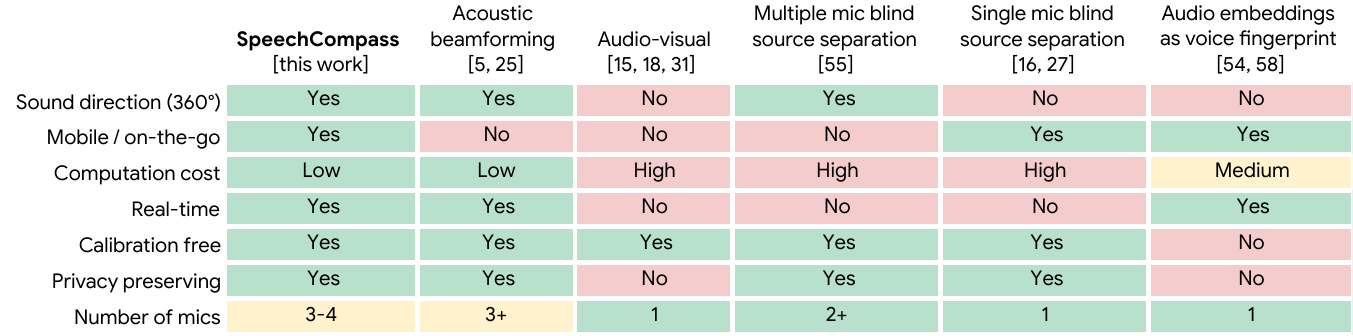}
  
\end{table*}

\textit{Acoustic beamforming} relies on classical signal processing techniques such as beamforming and localization from multiple microphones~\cite{anguera2007acoustic} to separate and diarize speakers. The main challenge for speaker separation is estimating a beamformer for each speaker using localization and other cues. Recently, neural networks have been employed to successfully aid in beamforming~\cite{heymann2016neural, yang2024binaural}. However, acoustic beamforming has been mainly applied to meeting room scenarios with a static microphone array, while SpeechCompass uses a similar multimicrophone technology for localization, which is applicable to mobile ASR throughout a user's everyday life.

\textit{Blind source separation.}
This approach separates the speakers using a single microphone without additional cues. The technique is challenging for classical signal processing, but various machine-learning techniques have been successful~\cite{erdogan2015phase, isik2016single}. Blind speaker separation is effective when done offline on the entire audio file, providing the model with access to both future and past content. This is, however, not possible in a real-time causal system, as in this paper, since the future is not accessible, and the model only processes a limited amount of past information. These constraints for real-time casual systems make blind separation inapplicable for real-time transcription of conversations.% more challenging and limits its performance. 

Recently, multi-microphone approaches have been combined with speaker separation and diarization. In~\cite{10446934} and~\cite{10508438} Taherian et al. tackle the challenge of speaker separation in multi-speaker scenarios, focusing on conversational or meeting environments. They leverage multi-channel audio and deep learning models to improve separation accuracy to enhance the performance of downstream speech applications like ASR. In one project~\cite{10446934}, the authors leverage an end-to-end diarization system to identify individual speaker activity and then use this information to guide a multi-talker separation model. In another approach~\cite{10508438} a multi-input multi-output (MIMO) complex spectral mapping model allows for robust speaker localization, and is used to reduce speaker splitting errors. The complexity of the processing and non-causal components make these solutions unsuitable for real-time processing.

\textit{Voice fingerprint audio embeddings.}
This approach extracts unique speaker embeddings from a single microphone~\cite{wang2018speaker,snyder2019speaker} and uses them for diarization. Principal Component Analysis (PCA) or other unsupervised methods are typically done on the embeddings. The speaker embedding approach has been the primary go-to for real-time diarization since it can run causally. Adding multi-microphone data to speaker embeddings has improved diarization accuracy~\cite{snyder2019speaker}. A key disadvantage of speaker embeddings is its reliance on implicit or explicit speaker enrollment, as the initial number of speakers is unknown. Requiring every conversation partner or nearby speaker to explicitly register through a voice sample is particularly impractical in dynamic mobile scenarios. Furthermore, there are privacy concerns as speaker embeddings can be considered biometric information, and asking people to enroll would be in conflict with the discreetness that is often desired for accessibility aids.

\textit{Audio-visual.}
Another approach has been to process audio and video using a multimodal model to separate speakers~\cite{gebru2017audio,ephrat2018looking}. The camera feed can help infer the active speaker from facial and lip motion when correlated with audio. Researchers have proposed deep learning models for audiovisual speaker separation that operate in the time-frequency domain and use cross-attention for audiovisual fusion~\cite{10446297}. The audio-visual model, while outperforming an audio-only model, however does not leverage spatial information from multiple microphones. The non-causal nature of the separation model (bi-LSTM) makes it ill-suited for real-time applications. This approach works best for meeting rooms and post-processing of recordings, since for mobile ASR, users typically do not point a camera at their conversation partners. There is potential for such applications for head-worn cameras (e.g., in smart glasses), although such approaches would still be dependent on a suitable field of view, sufficient bandwidth, and computing to process video streams. Such approaches will also have power and thermal implications for embedded devices with limited battery size.

Several commercial off-the-shelf solutions exist for mobile speaker diarization. The Ava mobile application~\cite{ava}, for example, allows diarization by connecting each speaker's smartphone to a network. However, this solution requires every speaker to set up and use their phone, which adds setup overhead. Speaksee~\cite{speaksee} is another solution that utilizes clip-on microphones for each speaker, with each microphone exclusively picking up speech from its designated wearer. These microphones connect to a central hub that provides diarized transcripts. This solution, while effective, requires dedicated hardware, making it more appropriate for formal meeting room scenarios.  While our approach may be less accurate than these solutions, because we are not using a dedicated microphone for each speaker, it does not depend on instrumentation or preparation by conversation partners, which is a crucial advantage in real-life situations whether with friends or strangers. It may even be prohibitive in certain scenarios where conversations are very brief (e.g., watercooler conversation) or the person wants to be discreet with their hearing accessibility needs.

\subsection{Visualizing non-speech sounds for accessibility use cases}
Speech is not the only aspect of sound that has been transcribed. For example, sound event recognition and visual alerts are helpful in hearing accessibility~\cite{sound_awareness, HomeSound, sound_detector_app,dhruv_sound_events}. Expressive captions~\cite{expressive_captions}, for example, visually convey the speech's emotion by altering the text's rendering based on the detected emotion. Closest to this work, research has looked into displaying sound localization cues from a microphone array using head-worm displays, such as HoloLens~\cite{holo_sound} and Head-Mounted displays (HMD)~\cite{dhruv_google_glass_localization}. Another work integrates localization into a watch~\cite{kaneko2013light} and displays sound location with LEDs. These proof-of-concept research projects use off-the-shelf devices, which do not meet power consumption, ergonomics, or form factor requirements needed for all-day use in everyday lives. In contrast, we leverage mobile phones that users already carry and use daily. While we demonstrate a prototype in a phone case form factor, the embedded hardware could also be implemented in wearable devices, such as head-worn displays. 

Due to captioning's focus on speech, fine audio qualities and structures, such as rhythmicity, are lost. To communicate such qualities, researchers have proposed real-time visualizations such as audio spectrograms~\cite{greene1984recognition} or Stabilized Auditory Images (SAI)~\cite{lyon2017human}, a visualization grounded in models of the brain, or tactile cues~\cite{vhp}. However, these approaches not only require extensive user training but also do not convey sound direction or leverage spatial information. Our approach can, however, be combined with those existing methods to extend their expressiveness. 

\subsection{Sound localization strategies} 
%Sound localization technology and algorithms
Multiple research projects have investigated sound localization, which can be divided into active and passive localization. Active localization sends out a signal and listens to the response. For example, Cricket \cite{Cricket} showed an indoor localization system using ultrasound beacons and detectors around the room. Active localization does not apply to speech, which is the main target of this paper.
As used in this work, a passive localization system listens with an array of microphones. Example of passive localization includes gunshot localization~\cite{valenzise2007scream}, rendering of a sound source in virtual reality~\cite{vr_sound_localization}, robot navigation~\cite{liu2010continuous}, and people tracking~\cite{people_walking_audio_id}. The Time Difference of Arrival (TDOA) is a popular technique for localization, where measuring the arrival delay between synchronized microphone pairs, allows the use of simple geometry to estimate the angle of arrival. 

Multiple localization techniques have been proposed in the literature based on the popular and effective TDOA estimation method. Most are variations~\cite{lee2008maximum,do2007real} on  generalized cross-correlation~\cite{knapp1976generalized}, performed in the frequency domain. Other approaches have been demonstrated, such as localization with one microphone~\cite{saxena2009learning} by adding structures to the microphone that change the sound based on incoming angle, or using two thin wires that change resistance based on the direction of sound~\cite{Sound_loc_and_vis_device}. 

Recent high-end phones use two or three microphones, which allows limited localization on some phones~\cite{phone_localization, phone_localization2}. However, the microphone placement on these phones is typically optimized for phone calls and video capture, with placements that are less well suited for localization, as microphones on the front and back are blocked when the phone is held or placed on a table. 

Several development platforms offer localization. For example, MiniDSP~\cite{MiniDSP} makes eight and sixteen microphone devices for localization and beamforming. ODAS~\cite{grondin2019lightweight} is an open-source platform for microphone array localization. Those platforms are, however, not designed for portable, battery-powered devices, given that their digital signal processing (DSP) chips have high power consumption ($\sim$500 mA).
%3d visualization~\cite{wang2012evaluation}

Although mobile phone ASR has many users, previous work shows that diarization and sound localization still need to be addressed, especially to make it practical for conversations with multiple speakers. While machine learning has improved diarization, the approaches are still challenging to apply to mobile ASR due to their numerous limitations (See Table \ref{table:tech_comparison}). Traditional multi-microphone processing techniques provide promising strategies for mobile ASR, which are practical today as shown through our technical implementation and evaluations.

\jl{say more about how your work build on the prior work}

%\alex{Maybe add something concluding that wraps up related work and talks about our complementary approach}\artem{Good idea, added a paragraph}

\begin{figure*}
\centerline{\includegraphics[width=0.7\linewidth]{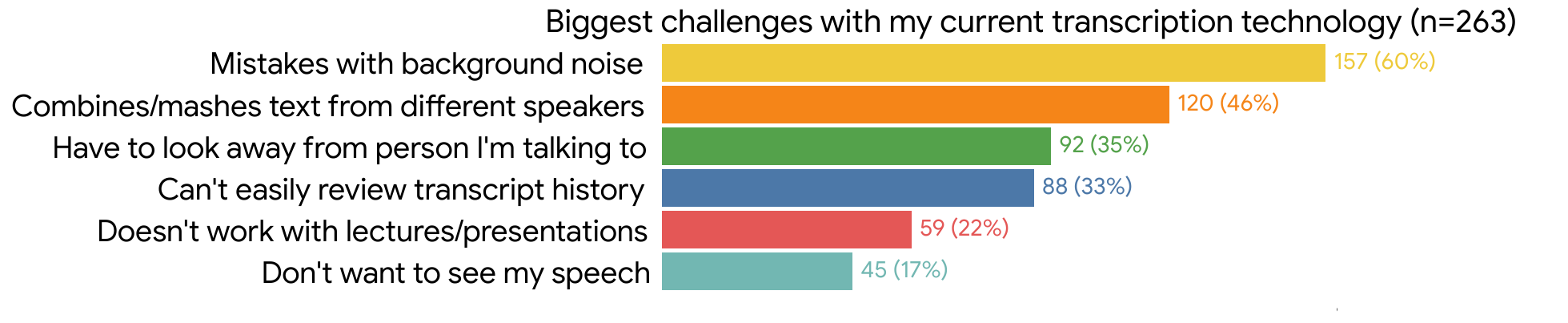}}
\caption{Participant responses to the question \emph{What are the biggest challenges with your current captioning or transcription device/technology? (select all that apply)?}}
\label{fig: survey-challenges}
\end{figure*}

\begin{figure*}
\centerline{\includegraphics[width=0.9\linewidth]{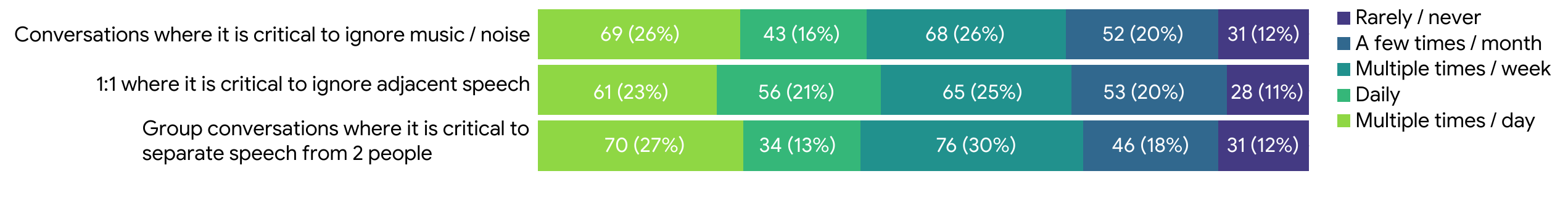}}
\caption{Survey results of how often participants encountered challenging scenarios with today's transcription technology. The number of participants and percentage is shown for each choice.}
\label{fig: survey-scenarios}
\end{figure*}

\section{Challenges with mobile captioning: Large-Scale survey with 263 frequent users} \label{surveys_foundational}
As we were interested in exploring the potential for more advanced mobile speech perception, we conducted a brief large-scale survey to learn about the challenges of using captioning for speech understanding in in-person meetings and conversations. % and opportunities for more advanced mobile speech perception. 

\subsection{Participants}
We used Google Surveys \cite{GoogleSurveys} to deploy a survey to the general population in the US of all ages and genders (\emph{“Android users of the Google Opinion Rewards app”}), screening for individuals that use technology to understand speech in meetings and conversations, and are frequent users of captioning technology. Our goal was to recruit deaf or hard-of-hearing participants, as we believed that they would have the most relevant experience and insights around mobile captioning technology and interfaces. To mitigate spam, the survey system analyzes question response times. By considering the distribution of response times across questions, it adapts to different question types and response patterns, rejecting sessions with unusually fast responses.

We acknowledge that our survey's focus on frequent users of captioning technology and the growing user base for mobile captioning limits its relevance for other populations, such as individuals who identify with Deaf culture and might be less likely to rely on ASR technology \cite{deaf_community_questionnaires}. Unfortunately, we cannot quantify the representation in the survey since restrictions from our institution do not allow us to collect participant hearing levels or use of sign language. 

Of the 1502 respondents that met our criteria, we focus on the 263 participants (18\%), who reported that they used captioning technology to understand people (not TV/video) multiple times per week or more frequently, and for 2 hours or more on the days that they used it. For these 263 participants, the platform reported that 49.8\% were women, 48.7\% men, and 1.5\% unknown, across all age ranges (27\%: 18--24, 33\%: 25--34, 15\%: 35--44, 12\%: 45--54, 6\%: 55--64, 7\%: 65+). 

The participants were prompted to select challenges among the choices from the list in Figure~\ref{fig: survey-challenges}. The choices were synthesized by user feedback from our previous experience with mobile captioning and informed by previous work in mobile captioning~\cite{localization_glasses, wearable_subtitle}

\subsection{Survey results: The use of captioning to understand people in conversations}
64\% of the participants reported daily use of captions in meetings or conversations to understand people, whereas 36\% used it multiple times per week. Half of the participants (49\%) use technology to understand people face-to-face for 2-3 hours on the day of use. Almost a quarter (23\%) of participants use captions for 4-5 hours on the days of use and another quarter (28\%) for 6 or more hours. Real-time captions, such as CART (69\%), and the \emph{Android Live Transcribe and Sound Notifications} app (55\%) were the top two technologies that were used daily to understand people.  %See Figure \ref{fig: survey-usage}.

% \begin{figure}
% \centerline{\includesvg[width=0.75\columnwidth]{images/survey/usage.svg}}
% \caption{How often do you use captions/transcriptions in meetings/conversations to understand people? (e.g., CART, Live Transcribe - not including closed captions for TV/video)}.
% \label{fig: survey-usage}
% \end{figure}

The top two issues with current transcription technology, as reported by our participants, were background noise (60\%) and the combining of text from different speakers (46\%), without the ability to separate them. Participants selected all that apply from the choices shown in Figure ~\ref{fig: survey-challenges}.

Finally, we asked participants about scenarios that are known to be challenging with today's transcription technology but have the potential to be addressed with more advanced microphone arrays and speech perception algorithms. Scenarios of interest included conversations where ignoring music, noise, or adjacent speech would be critical. We were also interested in group conversations and situations where separating speech from two people is critical. 68-70\% of participants experienced these scenarios multiple times per week or more frequently, whereas only 11-12\% rarely or never experienced them, as shown in Figure \ref{fig: survey-scenarios}.

% 38-48\% of participants experienced these scenarios daily, and 25-29\% experienced them multiple times per week. 

\subsection{Discussion}

Our large-scale survey enabled us to identify essential challenges with current transcription technology for frequent users of captioning and shows that 68-70\% of the participants are frequently in situations that today's technology cannot adequately support. The findings suggest that more advanced speech technology for suppressing noise from adjacent speakers, music, or noise could help address those issues and that speech separation technology can potentially improve group conversations through more readable transcripts. 

\section{SpeechCompass system} 
\begin{figure*}
  \centering
  \includegraphics[width=0.75\linewidth]{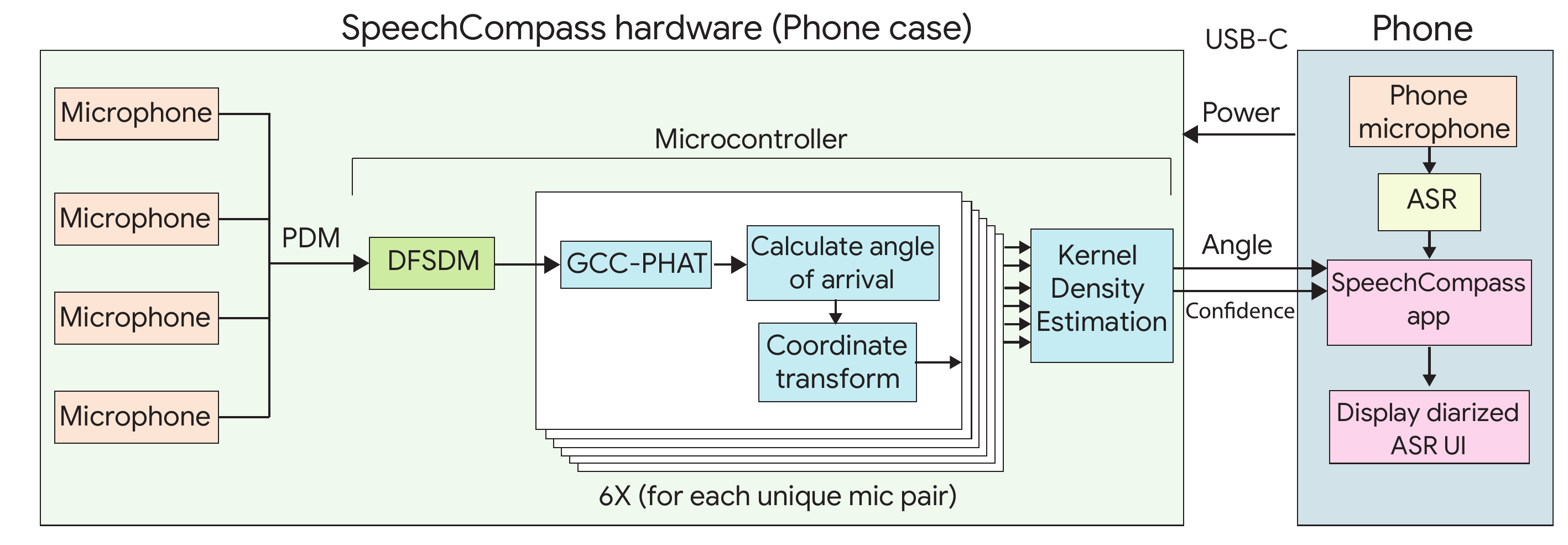}
  \caption{SpeechCompass system diagram. The phone case contains four microphones connected to a microcontroller. The audio localization algorithms run on the microcontroller, and the angle estimation is sent over USB to the phone. The SpeechCompass app combines ASR input and angle estimations to provide diarization and directional guidance for the mobile captioning UI.}
  \label{fig:system_diagram}  
\end{figure*}

%\section{SpeechCompass technology and implementation} 
This section details the SpeechCompass hardware, algorithms, and phone application. The SpeechCompass system diagram is shown in Figure~\ref{fig:system_diagram}.
\subsection{Design considerations}
Based on the foundational large-scale survey with 263 frequent users of mobile captioning (Section 3), previous work, 
and envisioned user journey, we outline the following design goals for the SpeechCompass prototype:
\begin{itemize}
    \item \textbf{All-day use.}
    The prototype should be low-power, ensuring that the battery could last a day or more, as 28\% of survey participants use captioning technology for six or more hours daily. Current DSP solutions~\cite{grondin2019lightweight, MiniDSP} draw about 500 mAh, which would require impractically large batteries, or significant impact on the phone battery.
    
    \item \textbf{Accurate 360\textdegree~localization.} The survey participants indicated that a major challenge is how transcription apps combine text from different speakers, which makes it difficult to use in group conversations. To address this challenge, localization accuracy should be under 20\textdegree~\cite{human_localization_error} to match human localization abilities and accurately distinguish different speakers. Also, 360\textdegree~azimuth localization is needed as it avoids front-back confusion, as illustrated in Figure~\ref{fig:tdoa_diagram}A. Also, since group conversations are dynamic, speakers and the phone may be located at any angle around the user.
    
    \item \textbf{Integration with existing phones and applications}. 55\% of the survey participants use mobile assistive apps, which is a strong motivator for us to target mobile phones, as they are already always with the user. The device should have multiple input and output options for audio and data, be designed to interface with external applications, and physically integrate with mobile phones. 
    
    \item\textbf{Low-latency, real-time processing.} To integrate with existing applications, the localization needs to be faster than real-time transcription latency, which is expected to be in the 30--300 ms range~\cite{yu2021fastemit}. Minimizing latency requires hardware-specific algorithm development and bypassing typical operating system latency. 
\end{itemize}

For someone who uses mobile captions daily, adding SpeechCompass for group conversations should be a low effort. The person would use the SpeechCompass phone case instead of a regular one. Their favorite captioning app would automatically use sound localization data to diarize the captions and separate the speakers. Considering the scenario where the person struggles to use mobile captioning around a dinner table, the text from different speakers would be uniquely presented (e.g., using colors or arrows) based on their position around the table. As phones add more microphones, SpeechCompass could become a pure software solution, utilizing multiple microphones on the phone. 

%\begin{figure}
%  \centering
%  \includegraphics[width=0.9\linewidth]{images/old_case.png}
%  \caption{Overview of the phone-case hardware with a physical interface. A) System diagram. B) Top and bottom of the PCB. C) One of the LEDs lights up to indicate the direction of the incoming sound. }
%  \label{fig:led_pcb}  
%\end{figure}

%\subsection{Multi-mic hardware for embedded sound perception}
\subsection{Multi-mic system for embedded perception}
We chose a 110 MHz M33 ARM Core (STM32L55, STMicroelectronics) as the main microcontroller. This processor provides the low-power and high computing capabilities needed for localization. \jl{consider moving the intro sentence about STM32 for hobbyists here if useful -- doesn't fit in into, but could make sense here}

Four digital microphones (MP34DT01-M, STMicroelectronics), which use the pulse density modulation (PDM) protocol, were arranged to have the largest distance between the microphones for finer resolution and support of 360\textdegree~of azimuth angles. Although three microphones could resolve similar angles, an additional microphone improves localization accuracy. Furthermore, when handheld, the four-microphone design is more robust to occlusions from the hand or fingers. The microphone data was collected at 16-bit resolution and 44.1kHz. The PDM to PCM (pulse-code modulation) conversion was done by the Digital Filter for Sigma-Delta Modulator (DFSDM) peripheral on the microcontroller. The same clock signal drove the four microphones, so they remained synchronized. We used flexible PCB-mounted microphones and a small main PCB, as shown in Figure~\ref{fig:pcb_design}B, C.

\rebuttaldelete{Two-channel audio output was provided with an audio codec chip (SGTL5000, NXP) with Digital to Analog Converters (DACs) routed to a 3.5 mm audio jack. The device includes two interface options: USB-C for data and multi-channel audio transfer, and analog audio input and output.}

%Two versions of the phone case and the PCBs were made. The architecture, shown in Figure~\ref{fig:led_pcb}A, and software remained the same. The first version included a physical interface, while the second version removed the physical interface to make the phone case compact.

%The first version is shown in Figure~\ref{fig:led_pcb}B
%The custom printed circuit board (PCB) size was 156$\times$73~mm, following the shape of a modern phone (Pixel 6, Google). The PCB was screwed to a custom-designed and 3D-printed (i3 MK3S+, Prusa) phone case. Eight tactile buttons were placed on the edges of the PCB. The buttons can be used for various quick user interface access. Also, twelve RGB LEDs (WS2812b, Neopixel) were positioned on the peripherals of the phone case. The LEDs were arranged to remain visible when looking at the phone screen or side of the phone, as shown in Figure~\ref{fig:led_pcb}C. 

%The second version was more compact and did not include an LED ring and buttons. It relied solely on displaying and controlling information on the phone screen. This version used flexible PCB-mounted microphones and a small main PCB, as shown in Figure~\ref{fig:pcb_design}B, C.

\subsection{Localization} 
\begin{figure*}
  \centering
  \includegraphics[width=0.70\linewidth]{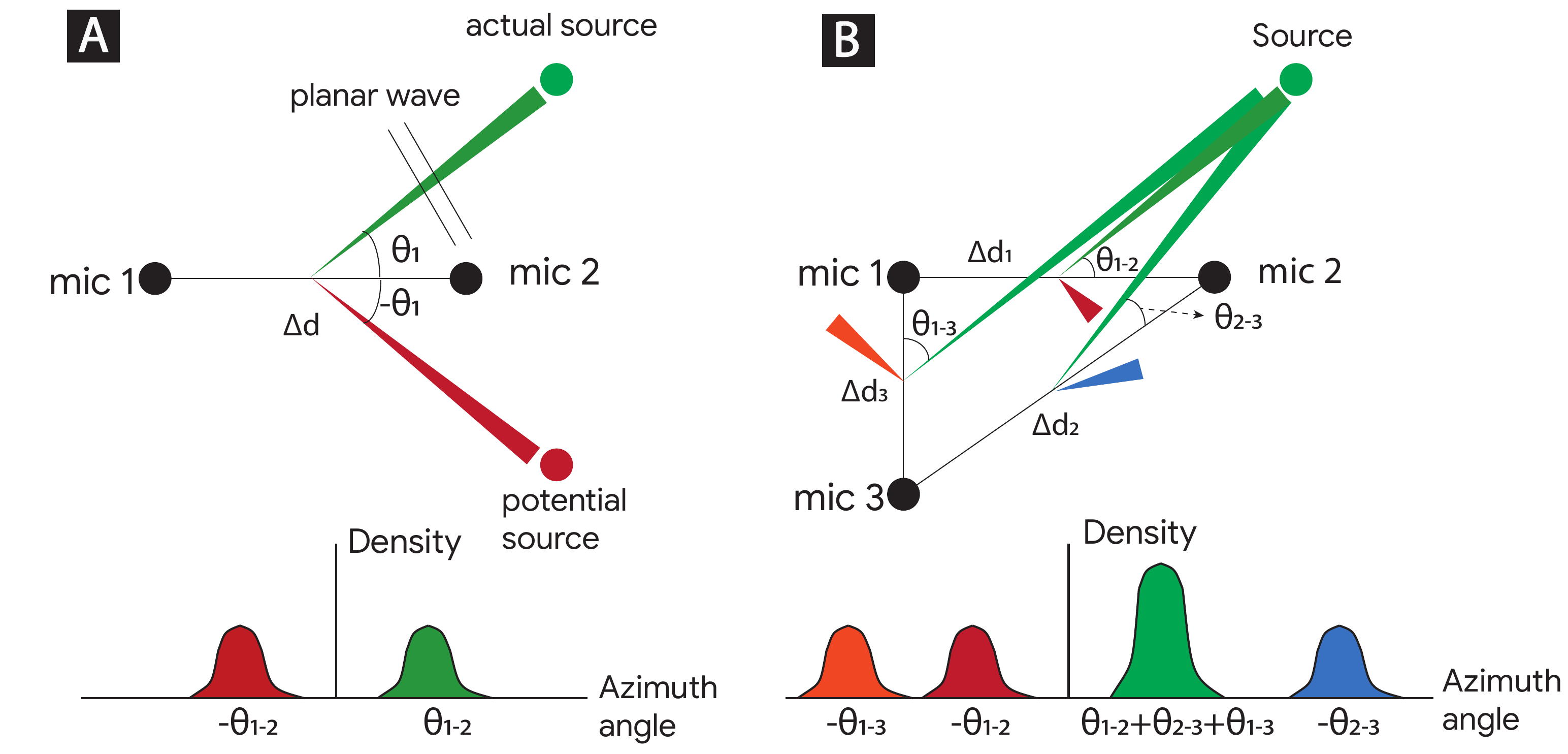}
  \caption{Visualization of localization methods with 2 and 3 microphone configurations. A) Localization with two microphones. The sound will arrive at microphone two before microphone one. This time difference could be used to estimate the angle of arrival. However, with two microphones, ambiguity exists, as the source could be at the inverse angle, shown as a "potential source." The graph on the bottom shows the kernel density estimation (KDE) with actual and potential sources. B) With three or more microphones, this angle ambiguity can be avoided. In our implementation, we use four microphones. The KDE from multiple microphone pairs will have the highest peak at the correct source.  }
  \label{fig:tdoa_diagram}  
\end{figure*}

The localization is computationally intensive, as it requires estimating the delay between all unique pairs of microphones. The delay estimation is usually done using cross-correlation. 

We implemented a variant of Generalized Cross-Correlation with Phase Transform (GCC-PHAT)~\cite{knapp1976generalized}. This approach is more robust to noise than standard cross-correlation and takes advantage of ARM's CMSIS libraries~\cite{cmsis_dsp}, which enable more efficient computations. Then, the time delay between the two microphones is extracted from the cross-correlation. Since the microphone geometry is known, the time delay is converted into the angle of sound arrival. The localization equations are provided in Appendix ~\ref{label:gcc_phat}.

\rebuttaldelete{The following equation was used:} 
\ifeq
\begin{equation}
\rebuttaldeleteeq{
G(f) = \mathcal{F}^{-1}\bigg(\frac{X_1(f)[X_2(f)]^*}{(|X_1(f)[X_2(f)]^*|)^{-0.3}}\bigg) ,
}
\end{equation}
\fi
\rebuttaldelete{
where $X_1(f)$ and $X_2(f)$ are the Fourier transforms of the two microphone signals, $[]^*$ denotes complex conjugate, and $\mathcal{F}^{-1}$ is the inverse Fourier transform. $G(f)$ is the resulting cross-correlation. 
We use partial normalization to the power of -0.3 since it provides more robustness to noise by giving less weight to delays around end-fires, which are more likely due to noise. Partial normalization deviates from the original GCC-PHAT as it uses full normalization to weigh all delays equally.
}
\rebuttaldelete{
The time delay between the two microphones is extracted from the cross-correlation in the following way:}
\ifeq
\begin{equation}
\rebuttaldeleteeq{
\Delta t = \frac{argmax(G(f))}{f_s} ,  
}
\end{equation}
\fi
\rebuttaldelete{
Where $f_s$ is the audio sampling frequency (44.1 kHz), and $argmax$ is the index of the maximum peak in the cross-correlation, corresponding to delay in samples. 
}
\rebuttaldelete{
To convert the time delay into azimuth angle (in the microphone plane), the microphone spacing needs to be known to calculate the maximum delay: 
}
\ifeq
\begin{equation}
\rebuttaldeleteeq{
\Delta t_{max} = \frac{\Delta d}{c} ,
}
\end{equation}
\fi
\rebuttaldelete{
where $c = 343 m/sec$ is the speed of sound, and $\Delta d$ is the distance between the microphones.} 

\rebuttaldelete{
Assuming far-field sound waves, we can use a simple formula to calculate the azimuth angle. The far-field approximation assumes planar sound waves and is valid approximately if the microphone is a meter or more away from the source.}
\ifeq
\begin{equation}
\rebuttaldeleteeq{
\theta_{azimuth} = cos^{-1}(\frac{\Delta t}{ \Delta t_{max}}) 
}   
\end{equation}
\fi

With just two microphones, there is front-back confusion. Different sources positioned at the inverse angle about the microphone pair axis will have exactly the same TDOA. (See Figure~\ref{fig:tdoa_diagram}A). 

%At least three microphones, which are not positioned on the same line, are needed to determine the 360-degree azimuth angle accurately. 
To accurately determine the 360\textdegree~azimuth angle, we need at least three microphones, positioned such that they span the largest plane possible (they cannot all be in-line). \alex{Tried to formulate this a bit clearer}

Considering the uncertainties of sound propagation, a statistical approach has to be used to determine the actual source location. We determine the TDOA for each microphone pair (six unique pairs for four microphones) and add a second potential TDOA. Then, we perform a coordinate transform so that each angle of arrival is aligned with global azimuth angles. Finally, we do Kernel Density Estimation (KDE) with a Gaussian kernel (bandwidth = 25) with the 600 latest samples. The highest peak will correspond to the angle of arrival, as demonstrated in Figure~\ref{fig:tdoa_diagram}. The KDE was wrapped around between 0 and 360\textdegree~to address the discontinuity around 0 and 360\textdegree. The localization is evaluated in Section~\ref{section:tech_eval}. 

%\subsection{SpeechCompass application: Mobile spatial speech perception} \label{subsection:app}
%\subsection{SpeechCompass application: Group conversation UI for ASR, leveraging localization and diarization } \label{subsection:app}
\subsection{SpeechCompass application: Mobile captioning with speaker separation} \label{subsection:app}

\begin{figure*}
  \centering
  \includegraphics[width=0.85\linewidth]{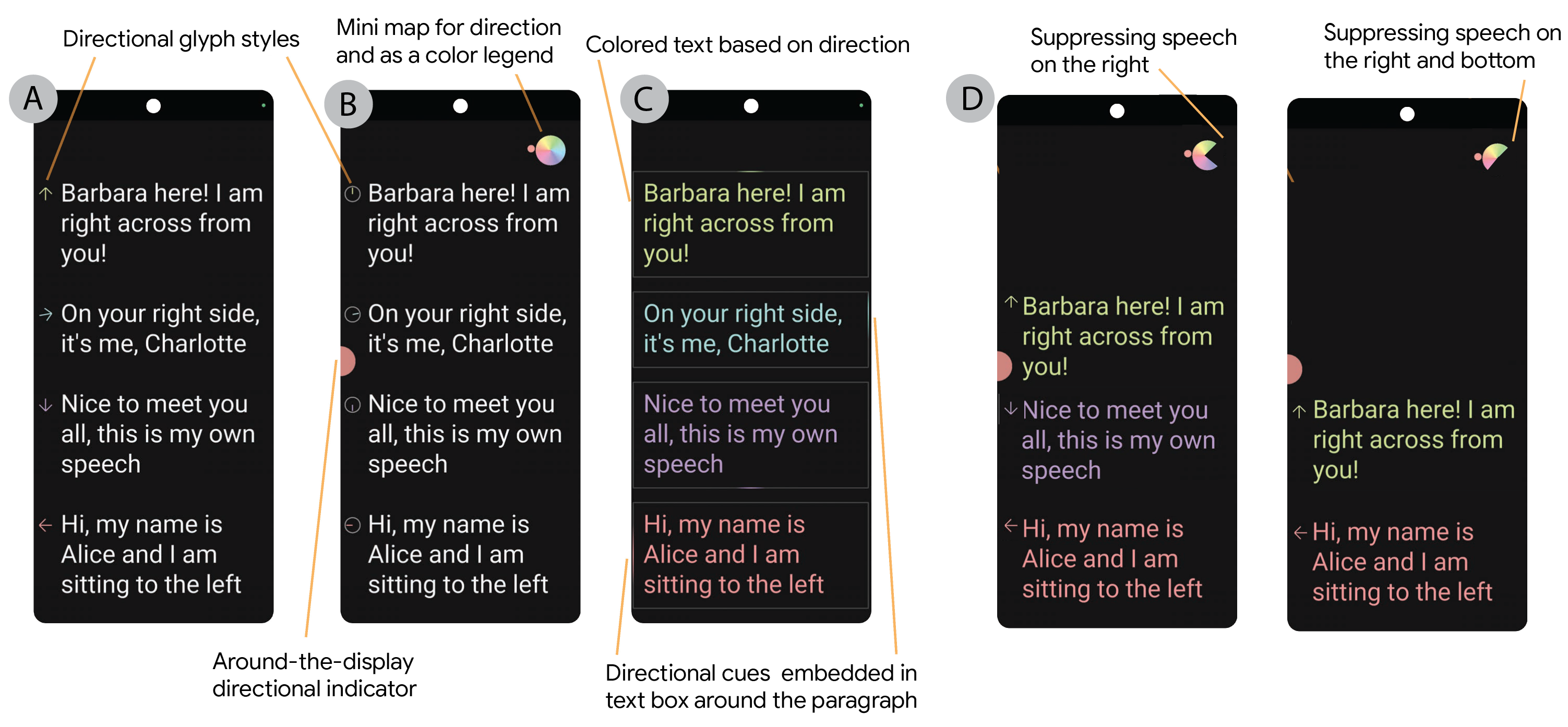}
  \caption{The mobile phone application with different direction visualization options. A) Directional glyphs are arrows next to the transcript, indicating the direction of speech. B) Minimap and directional glyph as radius inside the circle. C) Directional cues embedded in the color of the text and boxes around the text. D) Enabling speech suppression for right and bottom speech directions, as shown on the minimap. }
  \label{fig:phone_interfaces}  
\end{figure*}

To investigate the SpeechCompass technique's potential, we implemented a mobile ASR application (Android) with localization features, as shown in Figure~\ref{fig:phone_interfaces}. We used the USB Serial for Android library~\cite{AndroidUSBLibrary} to establish a data connection between the phone and the SpeechCompass microcontroller, which transmits localization data over USB-C. The application improves over existing single-source ASR applications in three ways, which we describe below.

\textbf{Real-time sound source visualization}. 
We use a semicircle overlay around the edges of the screen to indicate the current sound direction. The semicircle moves according to the azimuth angle of the audio, as shown in Figure~\ref{fig:phone_interfaces}B. (Alternatively, a vertical line could be used if only 180\textdegree~localization is available.) The radius of the semicircle is scaled according to localization confidence, corresponding to the KDE peak, with a larger semicircle corresponding to higher confidence. %We also implemented a loudness visualizer that computes the root-mean-square (RMS) of the current audio buffer and shows it as a scrolling waveform.

\textbf{Speaker diarization in transcription.}
APIs for mobile and web-based ASR have advanced significantly in recent years, but those APIs do not distinguish between multiple speakers or provide direction of the incoming speech. By combining SpeechCompass with mobile ASR, we enable legible transcripts by visually separating speech from different directions. 
We use the Android Speech Recognition API~\cite{AndroidSpeechRecognizer} to obtain real-time transcription and display the results on the mobile phone screen as vertically scrolling captions. The transcript is colored by directly mapping the sound arrival angle to the 360\textdegree color wheel, as shown in Figure~\ref{fig:phone_interfaces}C. In this mapping, the text is colored green if sound arrives from the top of the phone (around 90\textdegree) and red for bottom arrival (270\textdegree). Other direction indication options include showing a colored arrow next to the text or a directional glyph, as shown in Figure~\ref{fig:phone_interfaces}A, B. To accurately determine text color, we synchronize speech recognition and azimuth measurements by buffering azimuth angles after the \textit{onBeginningOfSpeech()} callback and stopping after the \textit{onEndOfSpeech()} callback. This way, only the angles detected during speech moments are analyzed, ignoring changes during silence or background noise. %The mode of the angles in the azimuth buffer was taken to be the correct angle.
We determine the resulting angle by computing the mode of the angles in the azimuth buffer.

\textbf{Speaker suppression.} 
As 60\% of the participants found mistakes with background noise to be an issue, we added unwanted speech suppression. With the speaker diarization, we can simulate speaker suppression, by letting the user hide speech from certain directions, or their own speech. Our implementation allows tapping on different edges of the screen. For example, as shown in Figure~\ref{fig:phone_interfaces}C, tapping on the right side of the screen will hide/show speech coming from the right.

\section{Technical Evaluation}\label{section:tech_eval}
In this section, we report on our technical evaluations of the main features of SpeechCompass. First, we characterize the underlying localization technology. We evaluate localization using an off-the-shelf mobile phone (Pixel 6, Google) and 360\textdegree~localization with the SpeechCompass hardware. Then, we determine the compute time for reliable localization, as it is critical for real-time diarization and visualization. Finally, for the real-world applicability of SpeechCompass, we characterize diarization accuracy in a realistic scenario where we also measure power consumption.

\subsection{Localization experiments setup}
By its nature, localization can not be done precisely due to time resolution. For example, with 80 mm microphone spacing and 44.1 kHz audio sampling rate, the maximum delay is $\pm$10 samples (0.23 ms). This configuration provides a resolution of 9$^\circ$. However, the actual accuracy is more nuanced, as it is obtained from multiple microphone pairs with different spacing and depends on many factors, such as the sound's frequency, environment, and loudness.
To better characterize localization accuracy, we conducted a series of experiments with the phone and SpeechCompass hardware. In the setup, the device under test was mounted on a rotating platform. A servo motor (XM430-W350-T, Dynamixel) was used to rotate the device around its axis (azimuth angle) and was synchronized with the data collection. 
A stationary speaker (S120, Logitech) was mounted 1.5 meters away to provide a direct sound path. 
Using the servo mount, the device was rotated by 10 degrees from 0--350$^{\circ}$ to obtain azimuth angle error. At each position, data was collected for 15 seconds. 
The loudness was calibrated with a reference sound meter (VL6708, VLIKE).
The speech audio came from the audiobook ``Alice's Adventure in Wonderland''~\cite{wonderland}. Rain~\cite{rain} was used as a realistic environmental sound with white noise properties. Each experimental condition was measured three times, and we report the mean value. 

\subsection{Localization: Off-the-shelf smartphone}
\begin{figure*}
  \centering
  \includegraphics[width=1.0\linewidth]{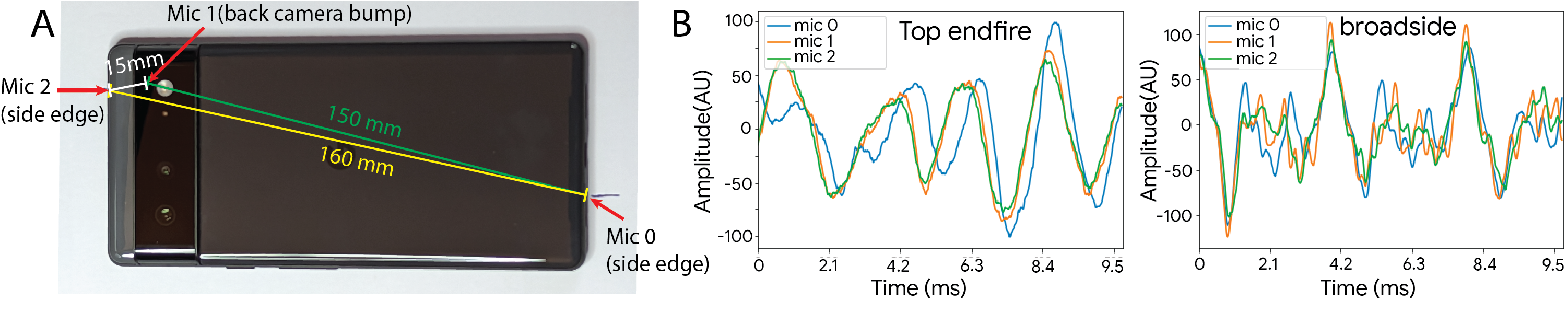}
  \caption{Localization on a mobile phone. A) Microphone positioning and distance. B) Examples of the raw microphone data from three microphones. The top graph shows the end-fire, with the bottom microphone facing the sound source, thus providing maximum delay between the microphones. The delay between mic 0 and mic 2 and 1 can be easily seen. The bottom graph shows broadside, where the microphones are the same distance from the sound, and thus have the same delay.}
  \label{fig:pixel_mics}  
\end{figure*}

Many smartphones today have two or three microphones, which has the potential to enable some localization. We investigate the feasibility and accuracy of localization with an unmodified, off-the-shelf phone (Pixel 6, Google). This phone has three microphones placed at the bottom, top and near the rear camera, as shown in Figure~\ref{fig:pixel_mics}. The top and bottom microphones are on the side edge of the phone, while the rear camera microphone is on the backside.

To characterize the opportunity and limitations, we ported our localization algorithm to Java Native Interface (JNI), so it could run on the Android platform. The two-mic-localization runs on the phone in real time.

With only a 15~mm separation between microphones 1 and 2, the delay between those two microphones was too small for useful localization. However, the other two pairs --- \{mic 0, mic 2\} and \{mic 0, mic 1\} --- provided useful localization data. %The geometry of the phone mics, we could only get 180-degree localization because of from-to-back confusion since mic 2 and mic 1 are too close to each other. 
Given this geometry, only 180-degree azimuth localization is possible for this device. If the camera microphone (mic 1) had been placed on the left side of the phone (with a larger baseline distance from mic 2), 360-degree localization would have been feasible. 

The mean error of environmental sound was 23.0 degrees, and speech was 16.0 degrees.  In comparison, humans also have up to 20-degree azimuth error when localizing sound without visual cues~\cite{human_localization_error}.
This demonstrates that localization on the phone can still be useful in limited use cases, for example, if the user does not expect any sounds from behind. We also observed that with such a limited amount of data, a simple histogram performed better than the Gaussian distribution approach, as there are not enough statistics for a reliable distribution. 
\alex{Seems like this could also be used for "Samsung interview mode" in portrait with one other person, or for 2 speakers in front with phone in landscape?}

% \subsection{Localization: SpeechCompass 4-mic hardware} 
\subsection{Localization: SpeechCompass 4-mic system} 

We varied two variables; type of sound (speech and environmental), and loudness in sound pressure level (SPL). 
The loudness varied from 50--65 dB SPL in 5 dB increments, where 0 dB corresponds to the human hearing threshold. This range covered the span of loudness typically experienced in daily conditions~\cite{db_chart}, from a quiet office (50 dB) to a normal conversation (65 dB). The ambient loudness in the experiment's room was 45 dB, so experimenting at this loudness or below was not practical. 

%Lastly, we played a mix of speech and environmental sound with x dB difference on an ortogonal speaker. This demonstrated performance of speech detection with presense of noise. 

\begin{figure*}
  \centering
  \includegraphics[width=0.7\linewidth]{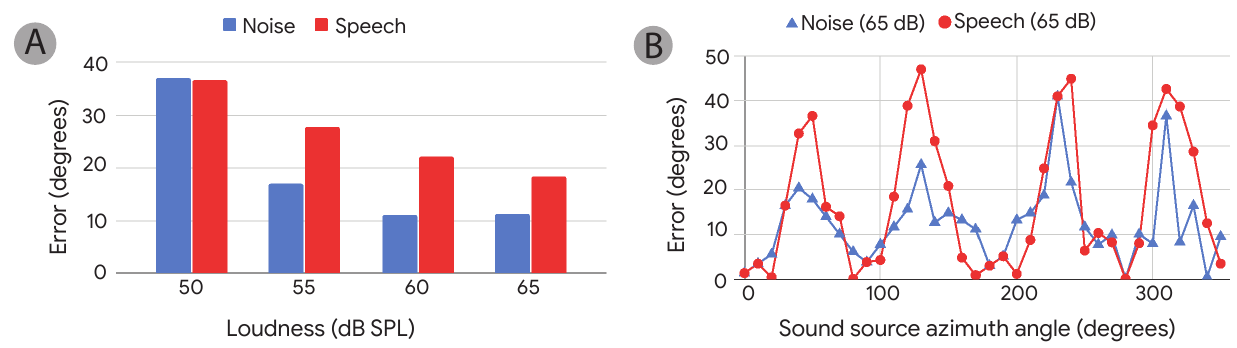}
  \caption{A) SpeechCompass azimuth angle measurement errors for speech and noise sound at different loudness levels. A mean of 0--350\textdegree~azimuth measurements is reported. B) Azimuth localization error at different angles. Data from one noise and speech trail are shown. The multi-mic device was rotated around its axis from 0--350\textdegree~while the source was stationary. A 15-second window was used for each estimate. }
  \label{fig:error}  
\end{figure*}

The experimental results show that sound could be localized reliably for both speech and environmental sound, with different loudness, and with 360$^{\circ}$ coverage. The overall error is reported in Figure~\ref{fig:error}A. The error was 11.1--22.1\textdegree~for normal conversational loudness (60--65 dB). In comparison, humans also have up to 20-degree errors when localizing sound~\cite{human_localization_error}. The error increased significantly as the sound got quieter. The GCC-PHAT picks up the loudest sound, thus making it difficult to estimate TDOA if the sound source is under or at the environmental noise level. 

\textit{Directional effect.}
The localization error has some dependence on the source angle, with an overall sinusoidal trend, as can be seen in Figure~\ref{fig:error}B. The lowest error was around 0, 90, 180, and 270\textdegree, whereas the highest error was observed at 45 and 135\textdegree. This sensitivity can be explained by the rectangular microphone geometry. With the current coordinate system, an angle such as 90\textdegree~results in 4 out of 6 microphone pairs in the end-fire and/or broadside in relation to the sound source. In such a configuration, the maximum and minimum delays are obtained, thus easier to measure. 

\textit{Type of sound.}
The error was lower for the environmental sound (11.3$^{\circ}$ at 65 dB) when compared to speech (18.3$^{\circ}$ at 65 dB). There are a few reasons why speech appears to be harder to localize reliably. First, speech has a very complex acoustic pattern in comparison to noise. There are many frequencies, sudden stops, and loudness variations~\cite{wilder1975articulatory}. This results in more reverberations and reflections, which can increase the error. Also, speech contains cyclic signals, which have ambiguity when subjected to cross-correlation, especially for high frequency, and speech contains signals up to 4 kHz. 

\begin{figure}
  \centering
  \includegraphics[width=0.9\linewidth]{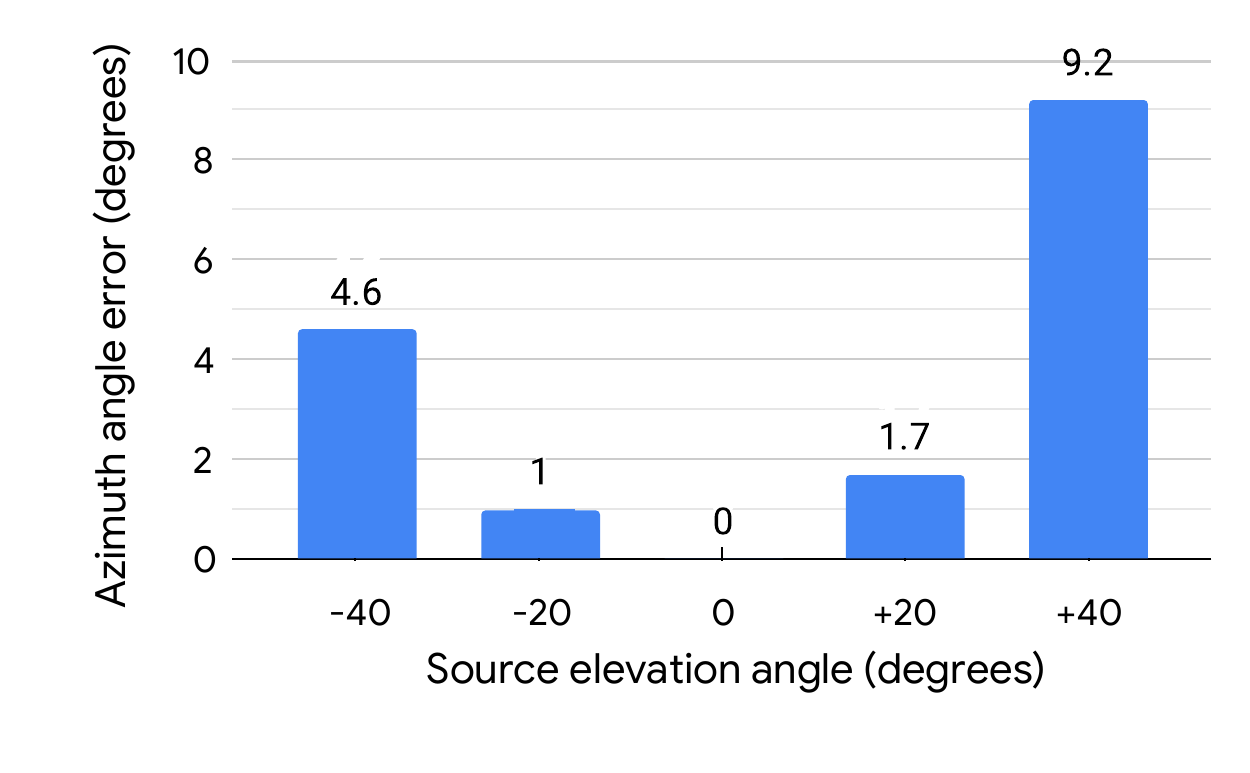}
  \caption{Effect of source elevation angle on the azimuth angle accuracy.}
  \label{fig:elevation_effects}  
\end{figure}

\textit{Effects of elevation angle.} In the above experiments, we test the azimuth angle error and keep the elevation angle at zero, so the audio source and microphone array are leveled. In practical use cases, the audio source might be located below or above the microphone array. For example, with the microphone array on the table, the speaker's head is usually above the array. To characterize the effect of elevation, we move the source elevation from -40 (source below mics) to +40\textdegree~(source above mics) in 20-degree increments, while keeping the azimuth angle at 90\textdegree. The distance is kept at 1.5 meters as in previous experiments and white noise is played at 65 dB. The results in Figure~\ref{fig:elevation_effects} show minor azimuth error (under 2 degrees) at -20 and +20\textdegree~elevation. The error becomes more pronounced (up to 9.2\textdegree) at -40 to +40\textdegree~elevation, as the source moves outside of the microphone 2-D plane. %The results show that calibration might be required to account for common use cases where speakers are usually above the microphone array. 
The 20\textdegree~elevation would result in a height of 0.55m, which would correspond to speakers sitting around a table with the phone 1.5 meter away. The 40\textdegree~elevation would result in a height of 1.26m, approximately corresponding to speakers standing around a table. The results therefore suggest that SpeechCompass could work well for seated conversation partners, but would need additional calibration to account for situations where speakers are further above the microphone array. \alex{Does this make sense? Please double check this reasoning!}

% 0.55 m (22'') - 20 deg, 1.26 m (50'') - 40 deg, 1.73 m (68'')- 60 deg

\subsection{Compute time to estimate direction} 
It takes 2.9 ms to compute the delay between two microphones on the 110 MHz microcontroller. Therefore the time to collect one data frame for six unique microphone pairs is 17.4 ms. This provides a lower limit on time to estimate direction. 

However, in practice, the estimation time could be much higher due to the uncertain nature of sounds like speech, as multiple measurements are needed for estimation. 
Therefore, we experimented to understand how quickly the device can react to the onset of sound in making an accurate direction estimate.

In a similar setup as the localization accuracy experiment, we positioned the device 1.5~meters from the sound source. As in the localization accuracy experiment, we used speech and noise, and measured them at different angles. 
 We collected data for 10 seconds without sound, then turned on the sound and collected the data for 50 seconds at 44.1 kHz. This measurement was conducted at angles from 0--360 in 60\textdegree~increments. 

\subsubsection{Results} 
We evaluated the number of samples it took for the sound source angle to converge around the actual angle by examining the KDE maximum after each additional sample. % (See Figure~\ref{fig:latency_example}).  

With the noise audio, localization required a mean of 12.7 samples (max: 19, min: 7). This number of samples was typically collected within one frame. So the latency was just the computational latency of 17.4 ms.

The speech audio needed on average 5.7X more samples to stabilize, with a mean of 72.7 samples (max: 149, min: 24). The speech contains more sparse localization information, and most measurements (80\%) did not have measurable microphone delays. As a result, a mean of 15 frames is required; thus, the mean estimation time is about 263 ms. However, estimation time can range approximately from 70--500 ms, depending on the speech characteristics.

\subsection{Power consumption} 
As the device is designed to be portable and powered by its own power source or the phone's battery, it is essential to understand its power consumption. The power consumption of the whole system was 28 mAh. The four microphones consumed 2.4~mA, while the microcontroller consumed 11.7~mA. The audio codec and other peripherals consumed the rest.
%\alex{was the power characterization running with the peripheral LEDs? or are these some status LEDs}\artem{I believe thats with the ring LEDs off in standby. }\alex{OK, let's remove it as that is confusing -- can be considered "other peripherals"}  
Therefore, the device could run speaker localization continuously for 18 hours with a 500 mAh battery. When powered by a smartphone, considering a representative smartphone's power consumption of 154 mA (3700 mAh battery for 24 hours), this represents an additional 18 percent of the power consumption. 

\subsection{Diarization accuracy with localization}
In this section, we evaluate diarization with the SpeechCompass hardware. A schematic view of the setup is shown in Figure~\ref{fig:dc_setup}A. 
We set up four speech sources at 0, 90, 180 and 270 degrees to mimic a turn-taking conversation with four talkers. The speech material is from Librispeech~\cite{panayotov2015librispeech}. A diffuse noise field is simulated using four loudspeakers simultaneously playing uncorrelated noise content. The noise types are babble noise (pub and cafeteria) as well as traffic noise. One synthesized conversation lasts an average of 40 seconds with each speaker speaking in turn (i.e., no overlap in the speech content). The noise level is set to create three signal-to-noise ratio (SNR) levels of 6, 12, and 18 dB respectively. A fourth scenario is generated in clean condition (i.e., no noise). For each scenario, 100 conversations are created, corresponding to about 1.2 hours of active speech.

\begin{figure*}
  \centering
  \includegraphics[width=0.6\linewidth]{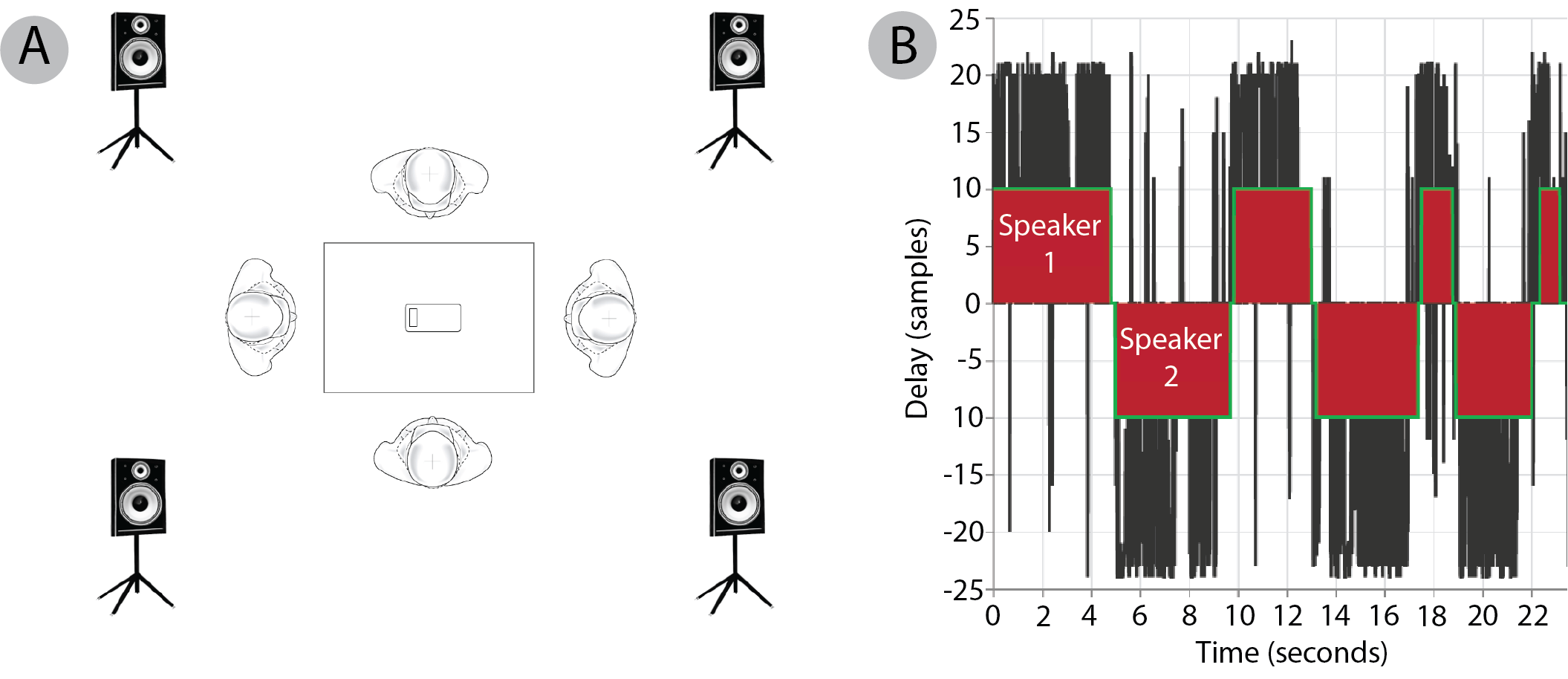}
  \caption{Diarization experiment. A) Data collection setup for diarization testing. The SpeechCompass phone case is placed on a small table in the center of the room. Head and torso simulators are used to play back speech material. Four loudspeakers are placed at the corners of the room to simulate a diffuse noise field. B) Example of TDOA data overlayed with diarization from two speakers for a conversation snippet. The extracted speaker labels from TDOA precisely follow the conversation turns. For TDOA, a frame size of 512 (11.6 ms) with no overlap at 44.1 KHz was used. Then, a running histogram of 522 ms was applied to get speaker labels.}
  \label{fig:dc_setup}  
\end{figure*}

An example diarized segment overlaying TDOA values is shown in Figure~\ref{fig:dc_setup}B, indicating the TDOA follows the speaker but still has some noise. 

In addition to the SNR sweep, we compared two microphone configurations using respectively three and four microphones, resulting in a total of eight experiments. We computed the diarization error rate (DER) for each scenario using the PyAnnotate toolkit~\cite{Bredin2020}. %DER is defined as the duration of speaker confusion error, false positives, and missed detections, divided by the total sesion time.\alex{Seems helpful with a definition of DER? I pulled in and adapted this, please verify and correct :)} \mathieu{would update to "
The DER is computed by summing the durations of false alarms, missed detections, and speaker confusions, and then dividing it by the total ground truth speech duration. The DER for the four-microphone configuration performed consistently better than the three-microphone one, across all four SNR conditions, with a relative DER improvement varying from 23 to 35\% for an average of 32\% as can be seen in Table~\ref{fig:dc_setup}. A four-microphone configuration allows for more unique microphone pairs (6) than a three-microphone configuration (3). This provides extra TDOA information that can be used by the diarization algorithm to segment and tag the speech content between the four talkers more accurately.

\begin{figure}
  \centering
  \includegraphics[width=1\linewidth]{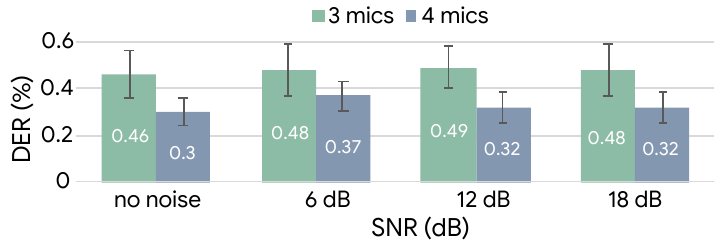}
  \caption{Diarization error rate (DER) for three and four-microphone configurations, across a sweep of SNR levels.\alex{I think caption should summarize the results/interpretation. Also, I swapped this table to a chart for easier comparison -- makes sense?}}
  \label{fig:dc_setup}  
\end{figure}

% Imagine of the diarization setup. 

% Where I got the current track from? 
%https://www.listenagainenglish.com/exercise-and-sports.html

%Calculate diarization error rate (DER). 
%https://github.com/wq2012/SimpleDER

% Looks like we can run puannotate directly: 
%https://colab.sandbox.google.com/github/pyannote/pyannote-audio/blob/develop/tutorials/intro.ipynb

\section{User evaluation with frequent users of mobile ASR: Lab study and online survey }
To evaluate the usability of our approach, we decided to conduct an in-person lab evaluation of the SpeechCompass phone case and the speech-to-text application (described in Section~\ref{subsection:app}), with frequent users of mobile transcription technology. We first conducted a large-scale online pilot study to inform the design of the in-person lab evaluation, which we conducted with eight deaf or hard-of-hearing participants, set up to mimic a realistic conversation scenario. 

\begin{figure*}
  \centering
  \includegraphics[width=0.75\linewidth]{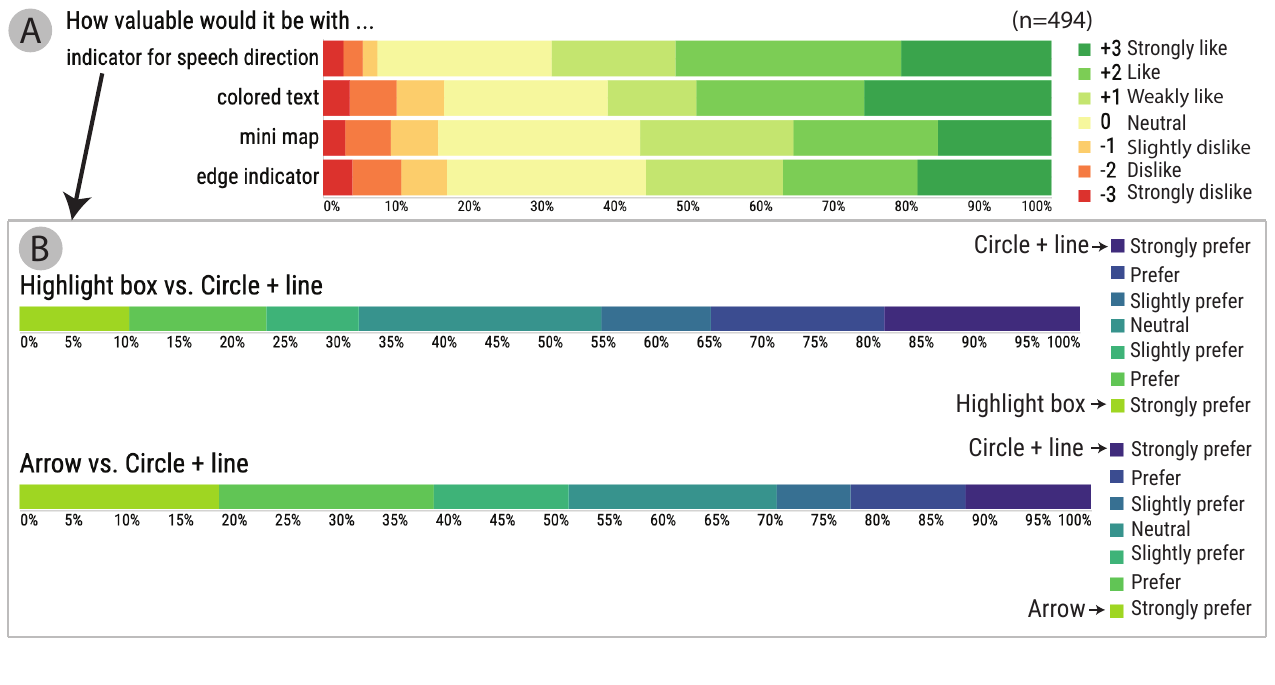}
  \caption{Participants' preferences for different visualization techniques in the online survey. A) Results indicating how valuable the specific indicator would be for the user. B) Preferences for the specific indicators for speech direction.} 
  \label{fig:user_preferences_online} 
\end{figure*}

\subsection{Large-scale, online survey (n=494)} In this survey, we use screenshots of our interactive UI prototypes to solicit initial user
feedback on the potential for our proposed approach, to guide the design of a more realistic in-person lab study.

The study was conducted using the same Google Surveys deployment and screening methodology as for the foundational study, detailed in Section 3. The participants were shown different UI renderings and were asked to rate them. The large-scale online survey could only show static images of the interfaces, due to limitations of the survey tool. Out of 985 respondents we focus our analysis on the 494 participants who use captioning technology multiple times per week or more frequently. 

As shown in Figure~\ref{fig:user_preferences_online}A, the colored text was found to be valuable by 60\% of participants. Glyph indicators for speech direction, which included arrow and circle+line indicators, were found valuable by 70\%. The Edge indicator and the mini map had a less positive reception. 

To better understand which glyph indicators were favored, we also asked targeted questions about them, as shown in Figure~\ref{fig:user_preferences_online}B. \emph{Circle + line} was preferred by 13.1\% more respondents than the \emph{highlight box} (45.1\% vs 32.0\%), and the \emph{arrow} was preferred by 21.9\% more respondents than the \emph{circle + line} (51.2\% vs 29.3\%).

\subsection{Lab study (n=8)}
\alex{explain and emphasize intention}
We recruited 8 participants from our institution who were frequent users of captioning technology. Five were female, three were male, and all were deaf or hard of hearing. One participant was 25--34 years old, four were 34--44, one was 45--53, and two were 65+ (we are only allowed to collect age ranges at our institution).

\begin{figure*}
  \centering
  \includegraphics[width=0.65\linewidth]{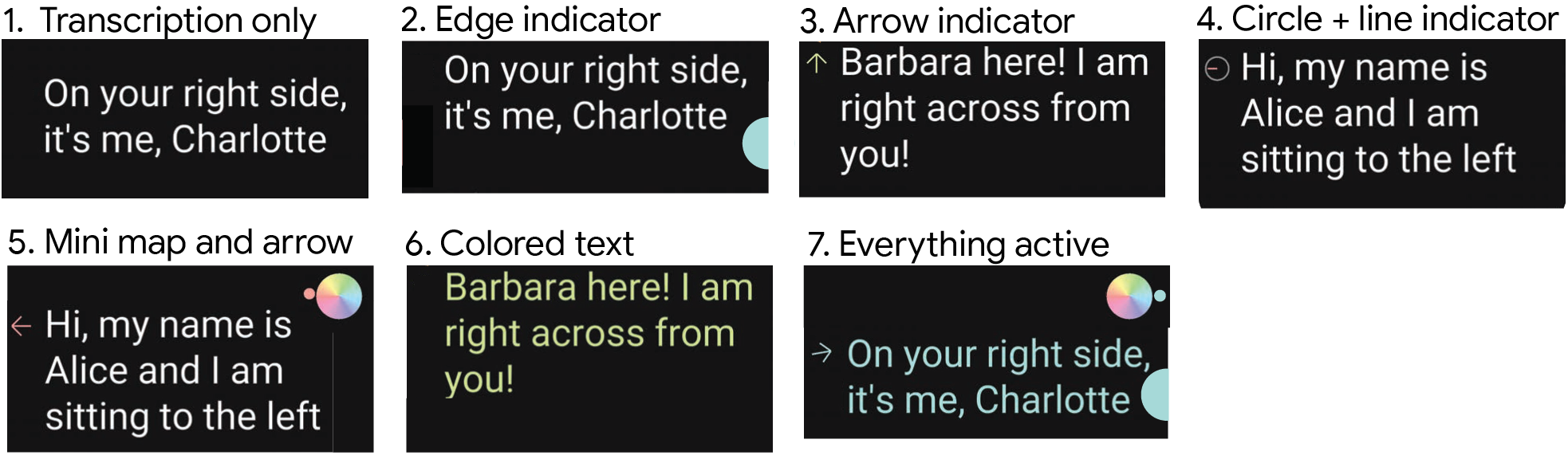}
  \caption{Examples of seven visualization scenarios that participants experienced in the in-person study.} 
  \label{fig:ui_options} 
\end{figure*}

% setup: https://docs.google.com/document/d/1akr5HVMgJb8Kd9KaEZJcdXn2S0IbHhd8JdBPTE0TiA0/edit?usp=sharing
The study took place in a quiet lab over approximately 60 minutes and used the phone-case prototype (Figure~\ref{fig:pcb_design}) with our mobile ASR application (Figure~\ref{fig:phone_interfaces}). First, the participant was introduced to the technology, prototype, and the purpose of the study. Then, the participant was asked to fill out a background survey, which included demographic questions and their current use and experienced challenges with transcription technology. Afterward, the participant was introduced to different visualization scenarios with the SpeechCompass application. The participant used the SpeechCompass transcription while sitting between the two experimenters, as they all sat around a small table with the SpeechCompass phone case in the center. In each of the seven conditions, which ran for 5 minutes, the experimenters sat across from each other and had short conversations about different topics. The participants were instructed to turn off hearing aid devices if they used any, and were asked to use the SoundCompass UI and transcript to follow the conversation. The experimenters' casual conversations included topics like weekend plans, hobbies, and the weather. The seven conditions, which used the ASR, diarization, and localization functionality for different visualization techniques, are shown in Figure~\ref{fig:ui_options} and presented with more UI context in Figure~\ref{fig:phone_interfaces}. The conditions were:
\begin{enumerate}
    \item \textbf{Transcription only}. The transcribed text is shown in white on a black background. 
    
    \item \textbf{Edge indicator}. A circle (``dot'') that moves around the edge of the screen to point to the currently active speaker. The color of the dot changes based on the direction. 
    
    \item \textbf{Arrow indicator}. A glyph using a colored arrow next to a white text block. The glyph points in the direction of the associated speech. 
    
    \item \textbf{Circle + line indicator}. A glyph using a circle with a directional line next to a white text block. The glyph points in the direction of the speech associated with the text. 
    
    \item \textbf{Mini map}. A colored circle with a smaller circle (``dot'') moves around its edge to point to the currently active speaker. The color of the dot changes based on the direction. 
    
    \item \textbf{Colored text}. The text is colored based on the direction that the associated speech was coming from. 
    
    \item \textbf{Everything on}. All indicators are turned on (except the Circle + line, as it couldn't be used simultaneously with the arrow). 
\end{enumerate}

%five isolated visualization techniques, baseline with just text transcription (no speaker information), and with all visualization turned one. Minimap was shown with an arrow, since we envisioned it would be combined with other techniques. 
After participants had completed all conditions, they filled out a form that asked them to rate how desirable each of the five visual indicator styles (\textit{Edge indicator}, \textit{Arrow}, \textit{Circle  + line}, \textit{Colored map}, and \textit{Colored text}) were on a 7-point Likert scale, from \emph{-3: Strongly dislike} to \emph{+3: Strongly like}. Finally, they were asked to rate the overall value of directional feedback to the transcription experience, how strongly they would recommend these features to users of mobile captioning, and whether they had any general free-form feedback about SpeechCompass. 

%After running the scenarios, participants filled out the second part of the survey, which asked them to rate each scenario and overall impression on a scale from -3:strongly dislike to +3:strongly like. Finally, the participants filled out free form feedback about the study. 

\begin{figure*}
  \centering
  \includegraphics[width=0.65\linewidth]{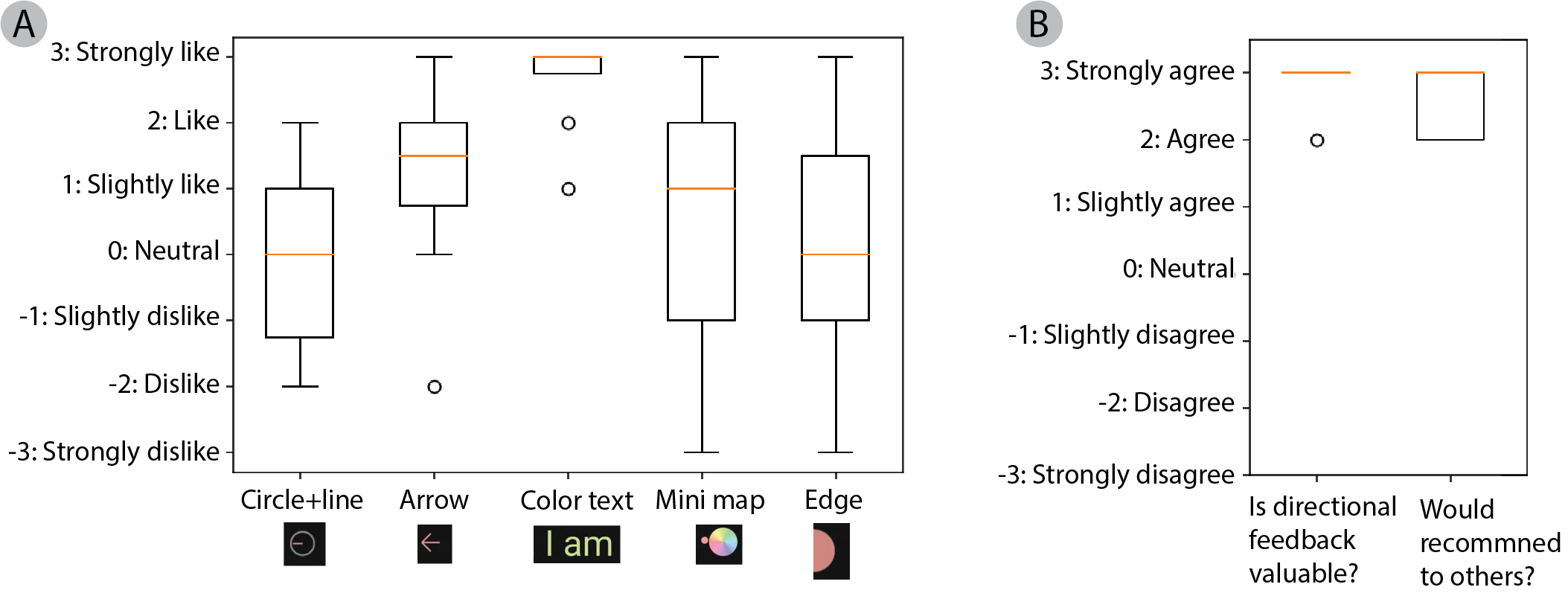}
  \caption{Boxplots of results of the in-person study. A) Participants' preferences for different visualization techniques. B) Overall opinions about augmented mobile ASR application.\alex{love these plots -- maybe to B you could also add the question about multi-people conversations as the leftmost, since it is also on same scale?} } 
  \label{fig:user_preferences} 
\end{figure*}

\subsection{Results}
Mobile transcription apps (e.g., Android Live Transcribe) were the most used communication technology for the participants. Specifically, three used them multiple times per day, one used them daily, three used them multiple times per week, and one used them rarely. 

75\% of participants frequently experienced the scenario where multiple people would get mixed up in the transcript (two multiple times per day, two daily, two multiple times per week). All participants agreed that it was challenging to participate in conversations when speech was combined from multiple people. 
%Similarly to the online survey, we asked participants to select the biggest challenges they experienced in their use of transcription technology (same options as in Figure~\ref{fig: survey-challenges}). where the majority (6/8) selected \textit{"Have to look away from the person I am talking to"}.  
\\

A Kruskal-Wallis (KW) test found a significant effect
on participant preferences for visualization techniques (P=.014).
The post-hoc pair-wise analyses using the Wilcoxon test with Bonferroni correction did, however, not show statistical significance between any pairs.
Of the five visual indicator styles that participants experienced, \emph{Colored text} was the most well-received (mean ($\bar{x})=2.625$), as it was rated positively by all the participants. %, with six strong like (+3), one like (+2), and one slight like (+1). 
The \emph{Arrow} indicator was also well-received ($\bar{x}=1.125$), with six positive, one negative, and one neutral participant.
%(one strong like (+3), three like (+2) and one slight like (+1)) and one dislike (-1) and one neutral (0)). 
Several participants noted that \emph{Arrow} and \emph{Colored text} worked well together: \emph{"Arrows + color seem to be most easier way to indicate the direction." (P2)} and \emph{"The combination of the colored text with the arrow was the most effective for me." (P7)}.

The other indicator styles received more mixed feedback. The feedback for both \emph{Edge indicator} ($\bar{x}=0.25$) and \emph{Circle + line} ($\bar{x}=-0.125$) was split between four negative and four positive participants. 
Some participants were concerned that \emph{Edge indicator} was distracting and not sufficiently discreet: \textit{"I do prefer the tool be as discrete as possible and would perhaps choose to avoid bright colored things moving around since this would be eye-catching and this kind of attention is often undesired" (P3)} and 
\textit{"Indicator moving around the edge was distracting and causing a bit of eye strain" (P2)}.
On the other hand, another participant found this style particularly useful: \textit{"the color dot moving to the speaker direction worked REALLY well" (P1)}. 
For \emph{Circle + line}, some participants struggled with its legibility: \textit{"If the analog direction indicators were larger (and translucent, or set behind)" (P8)} and \textit{"The lines in a circle were a bit slower and not as accurate (buggy)" (P5)}.
The \emph{Mini map} was rated positively by five participants and negatively by three. The most favorable participant stated: \emph{"this is also great for environmental awareness for those with single-sided hearing or no hearing at all." (P3)} and a participant who disliked the \emph{Edge indicator} commented: \emph{"steady map in the corner worked a bit better (P5)"}.

Overall, all participants agreed with the value of directional feedback ($\bar{x}=2.88$, seven Strongly agree:+3 and one Agree:+2) and would recommend these features to other users of captioning technology ($\bar{x}=2.63$, five Strongly agree:+3 and three Agree:+2): \textit{"I really liked that almost immediately I could tell that there was a speaker change, so that as soon as the text started to show up, I could better contextualize that text as attributed to a new speaker." (P1)}, \textit{"I'm very happy to see this tool being developed, it's a great addition to other speech recognition tools!" (P3)}, and \textit{"This prototype is definitely a life changer and I strongly believe that it will improve the quality of access to communication with speakers for many users" (P6)}.

\subsection{Discussion}
Consistent with the large-scale survey, the value of the diarization and localization features was immediate to all users. The participants were asked if directional guidance would be valuable in their mobile transcription experience. All eight users agreed. Also, all eight users would recommend this feature to mobile captioning users. 

While the large-scale survey helped inform our testing and exclude conditions (e.g., \emph{Highlight box}), the lab study allowed us to more rigorously evaluate the techniques in a realistic scenario. This difference became significant for the \emph{Edge indicator} and \emph{Mini map}, where issues, such as discreetness and distracting aspects, became evident during live usage. 

The results suggest that the combination of \textit{Colored text} and \textit{Arrow} would meet the preferences of most users, thanks to the balance of directional encoding and clarity. The arrow has redundant benefits too, since colored text might not always be reliably visible depending on lighting and screen conditions (e.g., strong sunlight, or dim display) and might also not be usable for colorblind users. The mixed feedback for other techniques indicates that the interface may also benefit from mechanisms that would allow users to customize the visualization style. Such customization could also apply to rendering properties, such as color, transparency, and line thickness, as some participants found \textit{Circle + line} particularly difficult to interpret. In both the large-scale survey and the in-person lab study, the \textit{Arrow} was preferred over \textit{Circle + line}. Through more customization options and extended usage in their daily lives, participants will be able to provide more nuanced feedback about these techniques.

\section{Limitations and future work}
In this work, the microphone array topology was designed for experimentation and integration with mobile phones. In the future, we plan to explore other form factors such as smart glasses and smartwatches, where multiple microphones could also be used. To do so, we plan to create a microphone array platform for rapid prototyping where microphone topology can be easily reconfigured and customized with a user-friendly UI and software library.

Localization needs to be accurate in everyday scenarios that may suffer from external noise and interference sources. We plan to add machine learning approaches to improve the noise robustness of our current localization, which is based on classical linear signal processing. 

In our user studies, errors due to background noise were identified as a major issue with mobile ASR. In future work, background noise could be suppressed by using a steerable beamformer, which would benefit from more microphones. In our experimentation with classical beamforming techniques like filter-and-sum beamformer~\cite{benesty2008conventional} and Minimum Variance Distortionless Response (MVDR)~\cite{xiao2017mvdr}, we achieved a few dB SNR (signal-to-noise ratio) improvement with our microphone geometry. With the recent advances in neural beamformers~\cite{li2016neural, yang2024binaural}, which are trained beamformers, higher SNRs could be possible. 

\alex{added back the longitudinal research}
\jl{go deeper with user studies with more people -- more design and evaluation with more people --> add more about the why, e.g., larger study to improve and find best visualizations (validate)}
We are interested in scaling up the user evaluation, both with a larger set of participants and over longer time periods and in ecologically valid settings.
We therefore plan to conduct longitudinal user evaluations with frequent users of mobile ASR to further advance our understanding of the usability of this approach in the wild.

\section{Conclusion}
Motivated by a foundational large-scale survey with 263 users frequent users of captioning technology, this work demonstrates how microphone array signal processing can supplement ASR by measuring and visualizing the spatial dimension of audio. Although ASR technology has greatly improved, diarization and localization features are not commonly available. 
To investigate the potential of enabling mobile captioning for group conversations, we implemented a low-latency 360\textdegree~localization algorithm that can run on general-purpose low-power microcontrollers, and a custom sound perception hardware solution with four microphones. 
%, in addition  and demonstrated that our localization algorithm can run on a mobile phone's existing built-in microphones, although limited in performance by microphone topology. 

Our technical evaluation of localization and diarization with the SpeechCompass microphone array demonstrates benefits over the traditional single-microphone configuration and pure machine-learning-based approaches. 

We introduced a mobile captioning app that uses sound localization to diarize and visualize speech directions for group conversations. 
%We enable communication of audio and localization data to the ASR application on the phone. 
The integration with our embedded software and hardware brought new capabilities to mobile ASR, including sound source location, speaker diarization, and user control of the diarization, allowing the suppression of specific speech directions. 
\jl{should probably mention the other large-scale study?}
Using the developed mobile phone application, we conducted an in-person study with frequent users of mobile ASR technology to gather feedback on the techniques and different visualization styles, and their potential for improving the captioning of group conversations. All the participants found the diarization, localization, and visualization features to be useful, and particularly appreciated the combination of a directional arrow and colored text. 

This work demonstrates that low-power microphone array processing can be integrated into new and existing mobile devices, thereby leveraging audio's natural spatial properties to enhance audio experiences and the understanding of speech. 

In the future, we hope that our approach will inspire the widespread adoption of advanced microphone arrays that natively unlock the potential of spatial sound processing and perception in mobile and wearable devices.

%\section{Acknowledgments}
\begin{acks}
We thank Sagar Salva and Dmitrii Votintcev for their ideas on prototypes and interaction designs. We are grateful to Pascal Getreuer, Richard Lyon, Alex Huang, Shao-Fu Shih, and Chet Gnegy for their help with algorithms. We also thank Shaun Kane, James Landay, Malcolm Slaney, and Meredith Morris for their feedback on this paper.  We appreciate the contributions of Carson Lau for the phone case mechanical design and Ngan Nguyen for electronics assembly. Finally, we thank Mei Lu, Don Barnett, Ryan Geraghty, and Sanjay Batra for UX research and design.
\end{acks}

\balance

% \balance
\bibliographystyle{ACM-Reference-Format}
\bibliography{references}
\appendix
\section{Appendix}

\subsection{Localization algorithm}  \label{label:gcc_phat}

The following GCC-PHAT formulation was used: 
\begin{equation}
G(f) = \mathcal{F}^{-1}\bigg(\frac{X_1(f)[X_2(f)]^*}{(|X_1(f)[X_2(f)]^*|)^{-0.3}}\bigg) ,
\end{equation}

where $X_1(f)$ and $X_2(f)$ are the Fourier transforms of the two microphone signals, $[]^*$ denotes complex conjugate, and $\mathcal{F}^{-1}$ is the inverse Fourier transform. $G(f)$ is the resulting cross-correlation. 
We use partial normalization to the power of -0.3 since it provides more robustness to noise by giving less weight to delays around end-fires, which are more likely due to noise. Partial normalization deviates from the original GCC-PHAT as it uses full normalization to weigh all delays equally.

The time delay between the two microphones is extracted from the cross-correlation in the following way: 
\begin{equation}
\Delta t = \frac{argmax(G(f))}{f_s} ,  
\end{equation}

Where $f_s$ is the audio sampling frequency (44.1 kHz), and $argmax$ is the index of the maximum peak in the cross-correlation, corresponding to delay in samples. 

To convert the time delay into azimuth angle (in the microphone plane), the microphone spacing needs to be known to calculate the maximum delay: 

\begin{equation}
\Delta t_{max} = \frac{\Delta d}{c} ,
\end{equation}

where $c = 343 m/sec$ is the speed of sound, and $\Delta d$ is the distance between the microphones. 

Assuming far-field sound waves, we can use a simple formula to calculate the azimuth angle. The far-field approximation assumes planar sound waves and is valid approximately if the microphone is a meter or more away from the source. 
\begin{equation}
\theta_{azimuth} = cos^{-1}(\frac{\Delta t}{ \Delta t_{max}})    
\end{equation}
\end{document}